\documentclass{article}
\usepackage[inner=3cm,outer=2cm]{geometry}
\usepackage{graphicx}
\usepackage{amssymb}
\usepackage[boxruled,linesnumbered]{algorithm2e}

\usepackage{amsmath}
\usepackage{relsize}
\usepackage{hyperref}

\usepackage{hyperref}
\hypersetup{colorlinks,
linkcolor=black,
citecolor=blue,
urlcolor=black
}

\newtheorem{thm}{Theorem}[section]

\newtheorem{prop}[thm]{Proposition}
\newtheorem{lem}[thm]{Lemma}

\newcommand{\RR}{\mathbb{R}}

\newcommand{\opmedian}{\operatorname{median}}
\newcommand{\oppvalue}{\operatorname{p-value}}
\newcommand{\oplogit}{\operatorname{logit}}

\newcommand{\eqd}{\stackrel{d}{=}}
\newcommand{\stge}{\ge_{\text{st}}}

\newenvironment{dfn}{\refstepcounter{thm} \bigskip \par \noindent
    {\bf Definition \thethm}\ }{\bigskip \par}
\newenvironment{exa}{\refstepcounter{thm} \bigskip \par \noindent
        {\bf Example \thethm}\ }{\bigskip \par}
\newenvironment{pf}{\medskip \par \noindent {\it Proof}\ }{\hfill
        \P \bigskip \par}

\begin{document}
\title{Exact confidence intervals for the mixing distribution from binomial mixture distribution samples}

\author{Pallavi Basu, Barak Brill, Daniel Yekutieli}

\baselineskip=18pt

\maketitle  

\begin{abstract}
We present methodology for constructing pointwise confidence intervals for the cumulative distribution function
and the quantiles of mixing distributions, on the unit interval, from binomial mixture distribution samples.
No assumptions are made on the shape of the mixing distribution.
The confidence intervals are constructed by inverting exact tests of composite null hypotheses regarding the mixing distribution.
Our method may be applied to any deconvolution approach that produces test statistics 
whose distribution is stochastically monotone for stochastic increase of the mixing distribution.
We propose a hierarchical Bayes approach, which uses finite Polya Trees for modelling the mixing distribution,
that provides stable and accurate deconvolution estimates without the need for additional tuning parameters.
Our main technical result establishes the stochastic monotonicity property of the test statistics produced by the hierarchical Bayes approach.
Leveraging the need for the stochastic monotonicity property, we explicitly derive the smallest asymptotic confidence intervals that may be 
constructed using our methodology.
Raising the question whether it is possible to construct smaller confidence intervals for the mixing distribution without making parametric assumptions on its shape.
\end{abstract}

\section{Introduction}

We present methodology for constructing exact pointwise confidence intervals for the cumulative distribution function (CDF) 
and the quantiles of the mixing distribution, $\tilde{\pi}(p)$, from a mixture distribution sample, $\vec{X} = (X_1, \cdots, X_K)$.
The underlying assumption in our work is that for $k = 1, \cdots, K$,  $X_k$ is independent $Binomial(m_k, P_k)$, for success probability $P_k$ 
drawn independently from $\tilde{\pi}(p)$. We denote this sampling model, $\vec{X} \sim \tilde{\pi} (p)$.
Our only assumption regarding $\tilde{\pi}(p)$ is that it has a CDF, $CDF_{\tilde{\pi}} (p)$,  at each $p \in [0,1]$.
At some points in the manuscript it will be more convenient for us to consider the logit of the success probability,
 $\theta = \oplogit(P)$,  in which case we will denote the mixing distribution on the logit scale $\pi(\theta)$.

\begin{figure}[htbp] 
\centering
\includegraphics[width=.9\textwidth,clip,trim={0 0 0 2cm}]{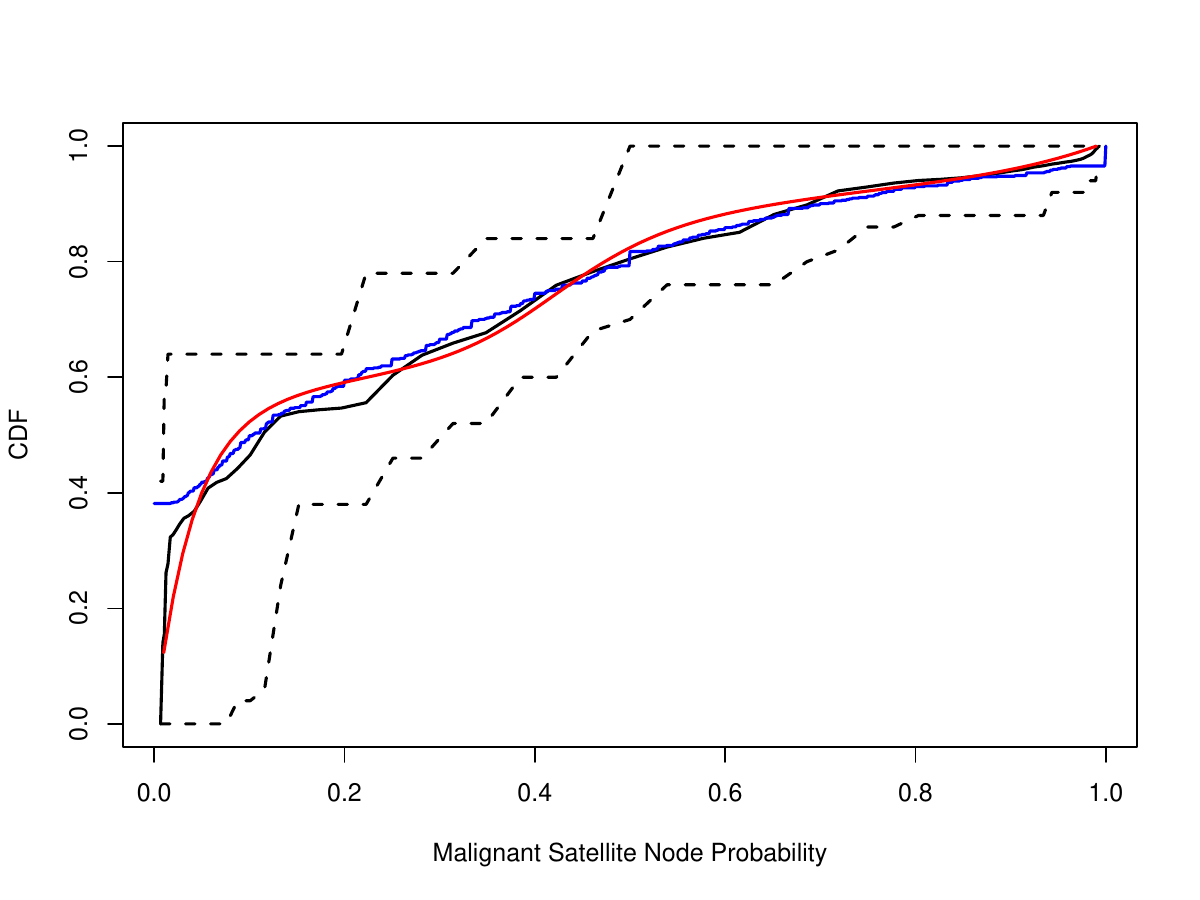}
\caption{Analysis of the intestinal surgery data example \cite{Ghol}.
The blue curve is the empirical CDF of the proportion of malignant satellite nodes.
The red curve is the estimate for the CDF produced by Efron's aproach.
The black curves are estimates and 90\% confidence curves for the CDF produced by our hierarchical Bayes approach.
}\label{figure:1_Efron_main_example}
\end{figure}

\bigskip
\begin{exa}   \label{exa0} Our motivating example is the intestinal surgery study \cite{Ghol} discussed in \cite{Ef16}.
The data for patient, $k = 1, \cdots, 844$, 
is  the number of satellite nodes removed in the surgery for later testing, in addition to the primary tumor, which we denote $m_k$, 
and the number of the satellite nodes that were found to be malignant, which we denote $X_k$.
\cite{Ef16} suggests a deconvolution approach for estimating the distribution of the fraction of malignant satellite nodes 
for the population of patients undergoing intestinal surgery.
In Figure~\ref{figure:1_Efron_main_example} we display the empirical CDF for the proportion of malignant satellite nodes, $\hat{P}_k = X_k / m_k$,
 the CDF estimates produced by Efron's approach, the CDF estimates and 90\% confidence curves produce by our hierarchical Bayes approach.
 
Intersecting the confidence curves with horizontal line, $CDF_{\tilde{\pi}} ( p) = q_0$, yields a  two-sided 90\% confidence interval for the $q_0$ quantile of $\tilde{\pi}$,
while intersecting the confidence curves with vertical line, $p = p_0$, yields a  two-sided 90\% confidence interval for $CDF_{\tilde{\pi}} ( p_0 )$.
The availability of confidence statements regarding the  distribution of the fraction of malignant satellite nodes is important for informed clinical decision-making.
For example, the two sided  90\% confidence interval for $CDF_{\tilde{\pi}}$ at $p = 0$ is $[0, 0.42]$. 
This implies that with probability $0.95$ in at most 42\% of the patients there is no malignancy in satellite nodes,
or alternatively, with probability $0.95$ in at least 58\% of the patients the malignancy has spread to satellite nodes.


\end{exa}

\medskip
In Section 2 we present the methodology  for constructing the confidence intervals.
The confidence intervals are acceptance regions for composite null hypotheses regarding the CDF of the mixing distribution.
The test statistics are estimates of the CDF of the mixing distribution.
We require the test statistic distribution to stochastically decrease with respect to stochastic increase of the mixing distribution.
Allowing us to perform exact tests that compare the observed test statistic value to the test statistic distribution for the worst-case null mixing distribution.
The confidence intervals we construct based on mixture distribution samples are 
considerably larger than the corresponding confidence intervals based on mixing distribution samples.
We show that the confidence intervals based on mixture distribution samples can be made smaller by incorporating a shift parameter
that  allow us to use the estimates of the CDF of the mixing distribution at smaller success probability values for testing hypotheses regarding the CDF of the mixing 
distribution at larger success probability values.
In Section 3 we introduce a hierarchical Bayes approach, in which the mixing distribution is generated by a finite Polya Tree model,
for estimating the mixing distribution from mixture distribution samples.
The main technical result of this paper is establishing stochastic monotonicity of the distribution of the mixing distribution estimates produced by the  hierarchical Bayes approach.
The hierarchical Bayes approach produces tighter estimates of the CDF mixing distribution, thereby producing smaller confidence intervals.
We provide a data driven procedure that uses a subset of the observations for determining the shift parameter value 
and uses the remaining observations for constructing the confidence interval for the mixing distribution.
 In Section 4 we discuss the asymptotic behaviour of the confidence intervals when the number of binomial samples tends to infinity
 and derive the smallest asymptotic confidence interval that may be constructed for a given quantile of a given mixing distribution.
 In Section 5 we discuss the scope of applicability of our methodology and the relevance of our asymptotic results to other mixture models.
 For improving the readability of the paper, we deferred the technical theoretical results, many of the algorithms, and additional analyses to Appendices,
and we illustrate implementation of all the methodology discussed in the text on the same simulated example.


\subsection{Related work}

There is extensive literature on the deconvolution problem,  sometimes called the measurement error problem, or the errors--in--variables problem
(see reviews \cite{De16}, \cite{Me09}, \cite{WW11}).
The deconvolution problem is typically phrased as estimating distribution $f_{\theta}$ from a sample $Y_1, \cdots, Y_K$,
with $Y_k = \theta_k + \varepsilon_k$, for $\theta_k \sim f_{\theta}$
and independent error term $\varepsilon_k$ generated from a known distribution. 
The classical approach (\cite{CH88}, \cite{SC90}) to the deconvolution problem involves  kernel density estimation that incorporates a Fourier inversion to accommodate for the noise. 
More recent work \cite{HL08} discusses the distribution and quantile estimators of and their  root--$K$ consistent optimality properties. 
Markedly \cite{Da16} follows up with a careful analysis of the convergence rates for quantile estimator. 
\cite{DH14} combines the knowledge of a parametric form of the unknown density along with the deconvolution kernel  estimator, 
through bias--correction or by the use of weights, yielding estimators of superior (say, when measured by integrated squared error) performances. 
\cite{Wa14} provides regularized optimization approaches using the marginal densities penalized by the $L_2$ norm in the former and by a predetermined order of derivative 
of the logarithm of the unknown density in the latter to encourage smoothness.
We are unaware of available implementable software for computing pointwise confidence intervals for the CDF or quantiles of the mixing distribution.

\cite{Saad} present methodology for constructing confidence intervals for the CDF of the latent mixing distribution 
 under the normal random-effects model for meta-analysis.
 Viewing the effect of an intervention in a new treatment group as independently sampled from the mixing distribution,
 they highlight the importance of the confidence intervals for the CDF for clinical decision-making that takes into consideration the heterogeneity 
of the intervention effect across treatment groups.


For completeness, we provide a succinct review of Bayesian approaches to the deconvolution problem. 
For instance, \cite{St08} fits a finite convex mixture of B--splines with a prior on the spline coefficients and applies to summaries of replicated observations.
The parameter distribution of interest is modeled as an infinite mixture of Gaussian distributions, an approach that is by now widely known as Dirichlet process mixture models 
(\cite{Fe73},  \cite{EW95}, \cite{Ne00}) in Bayesian nonparametrics. 
In Bayesian analysis deconvolution problems are sometimes viewed as empirical Bayes procedures, in which the prior distribution is estimated from the data.
\cite{Do18} offers an  intriguing empirical Bayes way to estimate the  hyperparameters with an emphasis on frequentist asymptotic properties as posterior contraction rates. 
In  \cite{Ef16},  which is the closest comparable work to ours,
the support of the mixing distribution is discretised on the logit scale to $(\theta_1, \cdots, \theta_N)$.
The mixing distribution at  $(\theta_1, \cdots, \theta_N)$ is modeled by a $r$-dimensional exponential family density  
$(\pi_1( \vec{\alpha}), \cdots, \pi_N( \vec{\alpha})) =   \exp \{Q \vec{\alpha} - \phi(\vec{\alpha})\}$, 
where $Q$ is a known $N \times r$ structure matrix, the parameter is $\vec{\alpha}  = ( \alpha_1, \cdots, \alpha_r)$ with $r << N$, 
and $\phi(\vec{\alpha})$ is a normalizing function.
For $k = 1, \cdots, K$ and $n = 1, \cdots, N$, let $P_{n, k}$ be the binomial probability of observation $\vec{X}_k$ for logit success probability $\theta_n$.
Thus the mixture distribution likelihood for $X_k$ is, $f_k ( \vec{\alpha}) = \sum_{n = 1}^N P_{n, k}  \pi_n ( \vec{\alpha})$.
The log--likelihood is further penalized to encourage shrinking in the parameter space,
$m(\vec{\alpha}) := \sum_{k = 1}^K \log( f_k ( \vec{\alpha}))  - c_0 \| \vec{\alpha}\|_2$,
where $c_0$ is a chosen constant and $\|\cdot\|_2$ denotes the Euclidean norm,
and $\vec{\alpha}$ is estimated by a maximization algorithm. 
Subsequently, once one realization of the $\widehat{\alpha}$ is obtained the bootstrap tool is applied to generate more datasets, 
 thus varying estimates of $\vec{\alpha}$ leading to the bias and the covariance calculations for the mixing distribution.
 Similarly, computations are performed  for the normal, and the Poisson variate and implemented in an R package \cite{EN16}.


\medskip
We approach the problem of constructing confidence interval for the mixing distribution
by the well--aged (Laplace, 1812 and Gauss, 1816)\footnote{Our understanding of these classics, written in French and German respectively,  
is from reading parts of the books \cite{LR05} (chapter 3) and \cite{Ha07} (chapter 8).}, 
but not pass\'{e} (applied, for example, in \cite{Mu95}; \cite{BW01}; \cite{Ba16}), 
idea of obtaining pointwise confidence intervals by inverting hypotheses tests. 
There is a skepticism\footnote{Read Andrew Gelman's 2013 blog for an engaging discussion.
 \url{http://andrewgelman.com/2013/06/24/why-it-doesnt-make-sense-in-general-to-form-confidence-intervals-by-inverting-hypothesis-tests/}} about the general idea. 
The concern is that when testing for parameter values if the parametric model is misspecified, or the data does not fit the family of distributions, 
the procedure may lead to a narrow confidence interval giving a false sense of precision. 
As the null hypotheses we test make no parametric assumptions regarding the mixing distribution,
the only model misspecification in our formulation is that the observed counts, $X_k$, are not independent $Binomial ( m_k, P_k)$.
We explicitly derive the smallest asymptotic confidence interval that may be constructed for quantiles of a given mixing distribution,
illustrating the dependence of the length of the confidence interval on the choice of quantile and on the shape of the mixing distribution.

The use of Polya trees for generating random distributions on dyadic partitions of the unit interval was introduced in \cite{Fe74}.
\cite{Lavine92} reviews the theoretical properties of Polya tree distributions, defines mixture of Polya trees and discusses choices of Polya tree parameters,
and suggests using mixtures of Polya trees as an alternative to parametric Bayesian analysis when the family of sampling densities is not known exactly.
\cite{Lavine94} suggests applying a Polya tree prior for mixing distribution for the nonparametric empirical Bayes problem.
\cite{Hans} presents a Metropolis-Hastings algorithm for obtaining  semi-parametric inference for mixtures of finite Polya tree models.
\cite{Walker97} uses Polya trees for specifying the distribution of the random-effects in hierarchical Generalized Linear Models,
 which correspond to extending our mixture model by incorporating covariate vector $\vec{Z}_k$ for each binomial count $X_k$, 
 $\oplogit(P_k) =   \theta_k + \vec{\beta}^T\vec{Z}_k$ for $\theta_k \sim \pi (\theta)$ and coefficient vector $\vec{\beta}$.
 \cite{Brill22} implements this modelling approach to microbiome analyses, 
 highlighting its usefulness for assessing non-pararametric random effects distributions and model coefficient testing and estimation.
A R software package implementing our hierarchical Bayes methodology for estimating and constructing confidence intervals for the mixing distribution 
is publicly available at \url{https://github.com/barakbri/mcleod}. 
The software supports computation of confidence intervals for mixing distribution quantiles or CDF values, as well as computation of confidence curves as shown in Figure~1,
it also provides mixing distribution and covariate estimates for non-parametric hierarchical Generalized Linear Models
for both binomial and poisson mixture samples.


\bigskip
\section{Confidence intervals for the mixing distribution}

One-sided $1 - \alpha$ confidence intervals for quantile $q_0 \in [0,1]$
of the mixing distribution and for the CDF of the mixing distribution at success probability $p_0 \in [0,1]$, are acceptance intervals for level $\alpha$ 
tests for two types of null hypotheses, either   $H^{LE}_0 ( q_0 ; p_0) :  q_0 \le CDF_{\tilde{\pi}} ( p_0)$ or  $H^{GE}_0 ( q_0 ; p_0)  :  q_0 \ge CDF_{\tilde{\pi}} (p_0)$.
Two-sided $1 - \alpha$ confidence intervals are derived by intersecting the corresponding left-tailed and right-tailed one-sided $1 - \alpha/2$ confidence intervals.
To simplify the presentation, in the manuscript we only consider the construction of left-tailed confidence intervals for quantiles of the mixing distribution by 
testing $H^{GE}_0 ( q_0 ; p_0)$.
The construction of the other types of confidence intervals is addressed in Appendix B.
For assessing the significance levels for the test statistic values, we will consider the worst-case mixing distribution 
$\tilde{\pi}^{min} (p ;  q_0, p_0)$, which assigns probability $q_0$ to the event $P_k = 0$  and probability $1 - q_0$ to the event  $P_k =  p_0$.

\begin{algorithm}[]
	\SetAlgoLined
	\SetKwInOut{Specify}{Specify}
	\SetKwInOut{Output}{Output}

	Specify decreasing sequence of success probabilities $1 > p_1 > p_2 > \dots > p_N > 0$
		
	\vspace{.15cm}
	
	\For{$n = 1, \cdots,  N$}{
		 Test null hypothesis $H^{GE}_0 ( q_0 ; p_n)$ at level $\alpha$ until the null hypothesis is first accepted.}	 

	If $H^{GE}_0 ( q_0 ; p_1)$ is accepted then set $p_{ul} = 1$ 

	If $H^{GE}_0 ( q_0 ; p_1)$ is rejected then $p_{ul}$ is the smallest $p_n$ for which  $H^{GE}_0 ( q_0 ; p_n)$ is rejected.
	
	\vspace{.15cm}
	
	Output confidence interval $[0, p_{ul}]$

	\caption{Algorithm for constructing left-tailed confidence intervals for the $q_0$  quantile of the mixing distribution}
	\label{algo1}
\end{algorithm}

\begin{prop}
The confidence interval $[0, p_{ul}]$ constructed in Algorithm \ref{algo1} is a valid $1 - \alpha$ confidence interval for  the $q_0$ quantile of $\tilde{\pi}$.
\end{prop}

\begin{pf}
The $q_0$  quantile of $\tilde{\pi}$ is either a success probability $p(q_0)$ such that 
$q_0 \le CDF_{\tilde{\pi}} \left( p (q_0) \right)$ and 
$ CDF_{\tilde{\pi}} (u) < q_0$ for all $u <  p(q_0)$,
or the subset of success probability values $\{ p : q_0 = CDF_{\tilde{\pi}} (p) \}$.
In any case, if $u$ is smaller than a $q_0$  quantile of $\tilde{\pi}$ then $CDF_{\tilde{\pi}} (u) \le q_0$.

Note that if $p_N$ is greater than all  $q_0$  quantiles of $\tilde{\pi}$, then any confidence interval constructed  in Algorithm \ref{algo1} 
covers all the $q_0$ quantiles of $\tilde{\pi}$ with probability $1$.
Otherwise, let $p_{crit}$ be the largest $p_n$ that is smaller than a $q_0$ quantile of $\tilde{\pi}(p)$. 
Thus, by construction $q_0 \ge CDF_{\tilde{\pi}} (p_{crit})$. 
Implying that $H^{GE}_0 (q_0 ; p_{crit})$ is a true null hypothesis
that is accepted with probability $\ge 1 - \alpha$. 
To complete the proof, notice that if $H^{GE}_0 (q_0 ; p_{crit})$ is accepted 
then the confidence interval constructed  in Algorithm \ref{algo1}  covers all  $q_0$ quantiles of $\tilde{\pi}(p)$.
\end{pf}


\begin{exa}   \label{exa1}
We apply Algorithm \ref{algo1} for constructing a  left-tailed 95\% confidence interval for the $0.40$ quantile of the mixing distribution
for simulated data consisting of $K = 80$ iid realizations,   $(P_1, X_1), \cdots, (P_{K}, X_{K})$, with $P_k \sim Beta(2,2)$ and  $X_k \sim Binomial (20, P_k)$.
For $n =  1 \cdots 99$, 
we set $p_n = (100 - n)/100$ and test  $H^{GE}_0 (0.40 ; p_n):   0.40 \ge CDF_{\tilde{\pi}} (p_n)$ with significance level $0.05$.
In this example, the test statistics are empirical CDF estimates for $CDF_{\tilde{\pi}}$,
we illustrate the use of a worst-case null mixing distribution to evaluate significance levels for the composite null hypotheses, 
and motivate the use of a shift parameter for deriving tighter confidence intervals. 

\medskip 
We begin by considering the ``no-noise'' case that we get to observe $\vec{P} = (P_1, \cdots, P_{K})$.
For this case, the test for  $H^{GE}_0 (0.40 ; p_n)$ is reject the null hypothesis for large values of  the empirical CDF of  $\vec{P}$ at $p_n$,
\begin{equation} \label{count1}
\overline{CDF} ( p_n ; \vec{P}) = \frac{ | \{ k : P_k \le   p_n \} |}{ K}.
\end{equation}
Note that for all mixing distributions for which $H^{GE}_0 (0.40 ; p_n)$ is true 
the distribution of the numerator in (\ref{count1}) is stochastically smaller than $Binomial ( 80, 0.40)$.
Let $Q_{0.95} ( 0.40)$ denote $0.95$ quantile of the $Binomial ( 80, 0.40)$ distribution. 
Thus the  test, reject  $H^{GE}_0 (0.40 ;  p_n)$ if  $\overline{CDF} ( p_n ; \vec{P}) > Q_{0.95}( 0.40) / 80$, has significance level $0.05$.
The $0.95$ quantile of the $Binomial(80,0.40)$ distribution is $Q_{0.95}  (0.40) = 39$, yielding critical value, $Q_{0.95}( 0.40) / 80 = 0.4875$.
In Figure 2 we display the $Beta(2,2)$ mixing distribution and the empirical CDF's of $\vec{P}$ and of $\vec{X}$.
The $0.40$ quantile of the $Beta(2,2)$ distribution is $0.433$.
The smallest $p_n$ for which  $H^{GE}_0 (0.40 ;  p_n)$  is rejected is, $p_{53} = 0.47$, for which  $\overline{CDF} ( 0.47 ; \vec{P}) = 0.4625$. 
Therefore the 95\% confidence interval for the $0.40$ quantile of $\tilde{\pi}$ based on $\vec{P}$ is $[0, 0.47]$.

\medskip
Now we assume that we only get to observe $\vec{X} = (X_1, \cdots, X_{80})$.
As before, we reject  $H^{GE}_0 (0.40 ; p_n)$ for sufficiently large CDF estimate values.
However now the CDF estimate is 
\begin{equation} \label{count2}
\overline{CDF} (p_n ; \vec{X}) = \frac{ | \{ k : \hat{P}_k \le  p_n \} |}{ K},
\end{equation}
for $\hat{P}_k = X_k  / 20$ and $K = 80$. It is easy to see that stochastically decreasing the mixing distribution stochastically increases the distribution of 
$\overline{CDF} (p_n ; \vec{X})$.
Thus $\tilde{\pi}^{min} (p ; 0.40, p_n)$,  which assigns probability $0.40$ to the event $P = 0$  and probability $0.60$ to the event  $P =  p_n$,
yields a test statistic distribution that is stochastically larger than the test statistic distribution of all null mixing distributions.
For  $P_k \sim \tilde{\pi}^{min} (p ; 0.40 , p_n)$,  the numerator of the test statistic in (\ref{count2}) is binomial with success probability $\gamma_0 (p_n)$, 
which we express
 \begin{eqnarray}
\gamma_0 (p_n) & =  & \Pr_{P_k \sim \tilde{\pi}^{min} (p ; 0.40, p_n )}  ( \hat{P}_k   \le  p_n)   \nonumber \\
& =  &    \Pr ( \hat{P}_k   \le p_n ,  \;  P_k = 0)  +   \Pr (\hat{P}_k   \le   p_n, \; P_k =  p_n )  \nonumber \\
& = &  \Pr ( \hat{P}_k  \le   p_n |   \;  P_k = 0) \cdot 0.4   +   \Pr (\hat{P}_k  \le  p_n | \; P_k =  p_n ) \cdot 0.6  \nonumber \\
& = &   0.4   +   \Pr ( \hat{P}_k  \le  p_n | \; P_k =  p_n ) \cdot 0.6. \label{prob:exp}
\end{eqnarray}
Where for $P_k = p_n$, $E \hat{P}_k = p_n$ and thus $\gamma_0 (p_n)  \approx 0.4 + 0.5 \cdot 0.6 = 0.70$.
Denoting the $0.95$ quantile of the  $Binomial(80, \gamma_0 )$ distribution by $Q_{0.95}  ( \gamma_0 )$, 
a significance level $0.05$ test for $H^{GE}_0 (0.40 ;  p_n)$ 
is reject the null hypothesis for $\overline{CDF} ( p_n ; \vec{X}) >  Q_{0.95}  (\gamma_0 (p_n)) / 80 $.
However as  $\gamma_0 (p_n) \approx  0.70$,
then $Q_{0.95}  (\gamma_0 (p_n))$ is much larger than $Q_{0.95} ( 0.40)$, and thus the resulting confidence interval is considerably longer.
 In the simulated example, $H^{GE}_0 (0.40 ; p_n)$  is last rejected 
 for  $p_{28} = 0.72$, for which  $ \gamma_0 (p_{28}) = 0.703$,  
 yielding  95\% confidence interval $[0, 0.72]$ for the $0.40$ quantile of $\tilde{\pi}$.
 
\medskip
Our solution for mitigating this problem is to specify a shift parameter $0 < \rho$,
for which we define 
\begin{equation} \label{exp:3}
p^*_n (p_n, \rho) = \oplogit^{-1} \left( \oplogit(p_n )  + \rho \right),
\end{equation}
for $n = 1, \cdots, N$ and the inverse-logit function $\oplogit^{-1}(\theta) = \exp(\theta) / (1+ \exp(\theta))$.
We then use the test statistic  $\overline{CDF} (p_n ; \vec{X})$ in (\ref{count2})
to test null hypothesis  $H^{GE}_0 (0.40 ; p^*_n)$, instead of $H^{GE}_0 (0.40 ; p_n)$.
For $P_k \sim \tilde{\pi}^{min} ( p ; 0.40,  p^*_n)$,  
the numerator of the test statistic in (\ref{count2}) is $Binomial(80, \gamma_0 (p_n, \rho))$, with
\begin{equation} \label{prob:3}
 \gamma_0 (p_n, \rho)  =   \Pr_{P_k \sim \tilde{\pi}^{min} (p ; 0.40, p^*_n )}  ( \hat{P}_k   \le  p_n)   = 
  0.4   +   \Pr ( \hat{P}_k \le  p_n | \; P_k =  p^*_n ) \cdot 0.6.
  \end{equation}
$\gamma_0 (p_n,  \rho)$ is decreasing in $\rho$.
For $\rho = 0$,  $\gamma_0 (p_n,  \rho) = \gamma_0 (p_n)$,
while for sufficiently large $\rho$, $\gamma_0 (p_n,  \rho) \approx 0.40$.
Allowing us to compare the same sequence of observed test statistic values,  $\overline{CDF} (p_n ; \vec{X})$ for $n = 1, \cdots, N$,
 to smaller critical values,  $Q_{0.95}  (\gamma_0 (p_n, \rho)) / 80$.
 The price paid for testing $H^{GE}_0 (0.40 ; p^*_n)$  instead of   $H^{GE}_0 (0.40 ; p_n)$,
 is that if the null hypotheses is first accepted for $p_{r+1}$,
 then the resulting confidence interval would be  $[0, p^*_{r} ]$, instead of the smaller confidence interval $[0, p_{r}]$.

 Setting $\rho = 0.5$, for $n = 1, \cdots, 99$,
 the null hypothesis $H^{GE}_0 (0.40 ; p^*_n)$ is last rejected for 
 $p_{45}  = 0.55$,
 with  $p^*_{45} ( 0.55, 0.5)  = 0.668$, $\gamma_0 ( 0.55 , 0.5)  = 0.512$,  $Q_{0.95} (0.512) = 48$ and  $\overline{CDF} (0.55 ; \vec{X}) = 50 / 80$.
Yielding  95\% confidence interval $[0, 0.668]$ for the $0.40$ the quantile of $\tilde{\pi}$.
The goal is to find large enough $\rho >0$, for which $\gamma_0 (p_n, \rho) $ is sufficiently close to $0.40$, 
 without inflating the length of confidence interval too much.
 Setting  $\rho = 1$ yielded the 95\%  confidence interval $[0,  0.731]$,
 while $\rho = 0.1$ yielded the 95\%  confidence interval $[0,  0.682]$.
Recall that without the shift parameter the 95\% confidence interval was $[0,  0.72]$.
\end{exa}

\begin{figure}[] 
\centering
\includegraphics[width=.8\textwidth]{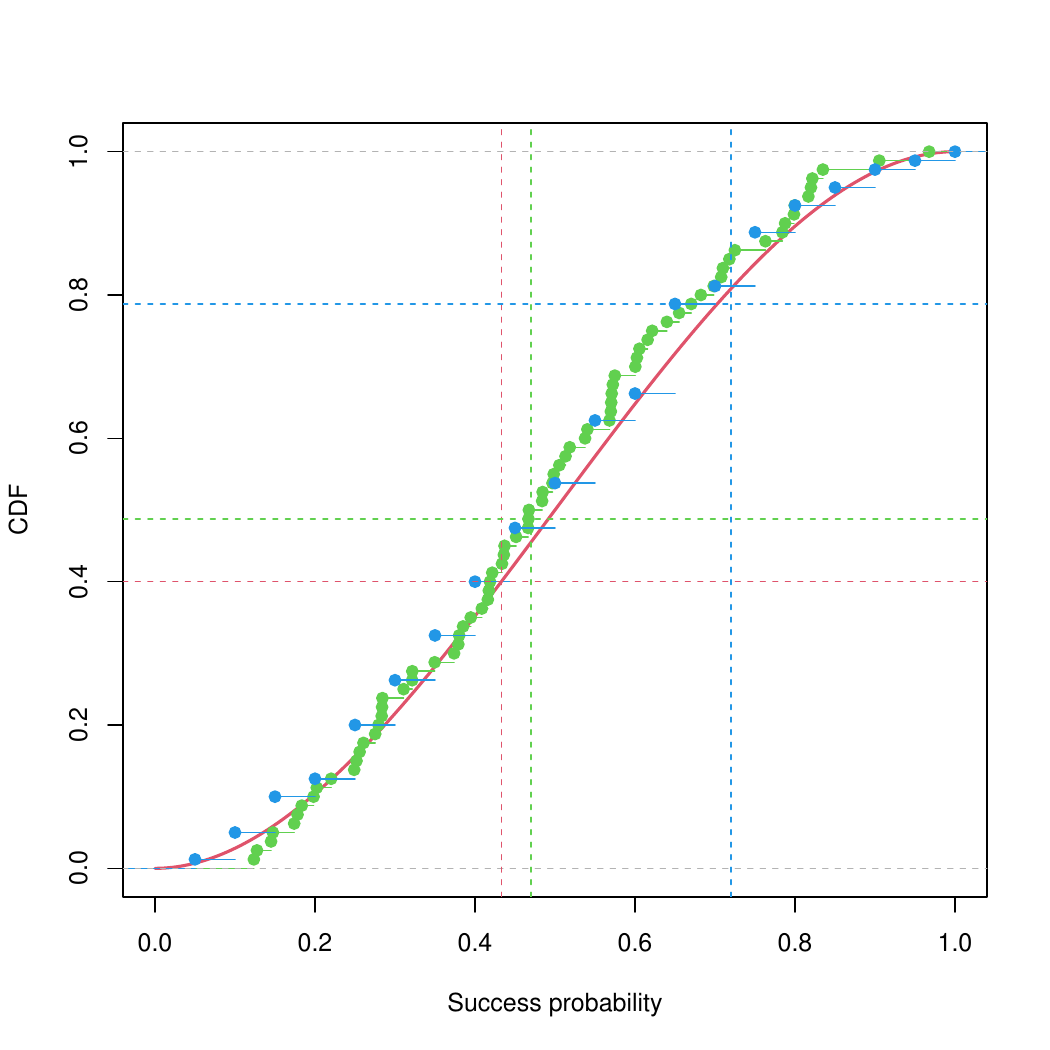}
\caption{Simulated example -- construction of left-tailed 95\% confidence interval for the $0.40$ quantile of the mixing distribution.
The red curve is the CDF of the $Beta(2,2)$ mixing distribution.
The green curve is the empirical CDF of $P_1, \cdots, P_K$, $P_k \sim Beta(2,2)$.
The blue  curve is the empirical CDF of $\hat{P}_1, \cdots, \hat{P}_K$, for $X_k \sim Binom(20, P_k)$.
The red, green and blue dashed horizontal lines curves are drawn at CDF values $0.4$, $0.4875$, $0.7875$.
The red, green and blue dashed vertical lines curves are drawn at success probability values $0.433$, $0.47$, $0.72$.
}
\end{figure}



\bigskip
\section{Hierarchical Bayes deconvolution-based tests}

In Example \ref{exa1}, we used the empirical CDF of $\vec{X}$ as the statistic for testing the null hypotheses.
In this section, we introduce methodology for estimating the mixing distribution, and use the estimates of $CDF_{\tilde{\pi}}$  for testing the null hypotheses.
As this method produces tighter estimates of $CDF_{\tilde{\pi}}$  it yields shorter confidence intervals for the same value of shift parameter,
while also making it possible to work with smaller shift parameter values.

Our estimate for $CDF_{\tilde{\pi}}$ is the posterior median of the CDF of the mixing distribution for a hierarchical Bayesian generative model for the 
distribution of the mixing distribution, the sequence of the success probabilities, and the sequence of observed binomial counts.
An important feature of this estimation method, in cases where the sample sizes, $m_k$, are different for each binomial sample,
 is that as the effect of each binomial count, $X_k$, is determined by its likelihood function,
the model propagates greater uncertainty regarding the value of $P_k$ for samples with smaller sample sizes $m_k$.

\subsection{Review of the finite Polya tree model}

In the hierarchical model we evaluate the mixing distribution on the logit scale. 
The basis for our hierarchical model is a finite Polya tree (FPT) model
that generates random mixing distributions with step function densities on a $L$ level dyadic partition of ${\cal I}_0  \subseteq \RR$,
specified by the endpoints vector $\vec{a} = (a_0 <  a_1 <  \ldots < a_{2^L})$.
${\cal I}_{0}  =  [ a_0,  a_{2^L}]$ with subintervals,  
${\cal I}_{l, j}  =  ( a_{ (j-1) \cdot 2^{L-l}} ,   a_{j \cdot 2^{L-l}}]$ for $l = 1 \cdots L$ and $j = 1 \cdots 2^l$.
 The parameters of the FPT model are Beta distribution hyper-parameters $( \alpha_{l, j}, \beta_{l, j})$,
 with $l = 1 \cdots L$ and $j = 1, \cdots, 2^{l-1}$.
 The  FPT model generates the following components.

\medskip
\noindent {\bf a. Independent Beta random variables}
 $ \phi_{l, j} \sim Beta ( \alpha_{l, j}, \beta_{l, j})$.
 The Beta random variables specify conditional subinterval probabilities:
 $\Pr({\cal I}_{1, 1} | {\cal I}_0  ) = \phi_{1, 1}$, $\Pr({\cal I}_{1,2}  | {\cal I}_0 )  = 1 - \phi_{1, 1}$,
and for $l = 2, \cdots, L$ and $j = 1, \cdots, 2^{l-1}$,
 $\Pr({\cal I}_{l, 2 \cdot j - 1} | {\cal I}_{l-1, j}) = \phi_{l, j}$ and $\Pr({\cal I}_{l, 2 \cdot j } | {\cal I}_{l-1, j}) = 1 - \phi_{l, j}$.

\medskip
\noindent {\bf b. Subinterval probabilities.}  The subinterval probabilities, $\Pr({\cal I}_{l,j})  = \pi_{l, j}$,
are products of the conditional subinterval probabilities.
$\pi_{1, 1} =  \phi_{1, 1}$ and $\pi_{1, 2} = 1 - \phi_{1, 1}$.
For $l = 2 \cdots L$ and for  $j = 1 \cdots 2^{l-1}$,
$\pi_{l,2 \cdot  j -1} = \phi_{l,  j} \cdot \pi_{l-1,j}$ and $\pi_{l,2 \cdot  j} = (1- \phi_{l,  j}) \cdot \pi_{l-1,j}$.

\medskip
\noindent {\bf c. Step function density function.}
The components of $\vec{\pi}_L = (\pi_{L, 1}, \cdots, \pi_{L, 2^L})$ specify a distribution with step function density,
\begin{equation} \label{def-step}
f ( \theta  |  \vec{\pi}_L  ; \vec{a} )  =  \pi_{L,1} \cdot  \frac{ I_{(a_0 , a_1]} ( \theta) }{a_1 - a_0} + 
\pi_{L,2} \cdot \frac{ I_{(a_{1} , a_2]} ( \theta) }{ a_2 - a_1} + \cdots + 
\pi_{L,2^L} \cdot \frac{ I_{(a_{2^L-1} , a_{2^L}]} ( \theta) }{ a_{2^L} - a_{2^L-1}},
\end{equation}
 for $\theta \in {\cal I}_0$ and indicator function $I_{(a_{j-1}, a_j]} (\theta)$.



\bigskip
\subsection{The hierarchical Bayes generative model}
\noindent For computing the deconvolution based test statistic we assume that the  step function density,
logit success probabilities, $\vec{\theta} = ( \theta_1, \cdots, \theta_K)$, 
and binomial counts, are generated by the following hierarchical model.

\begin{dfn} {\bf Generative Model} \label{def:seq} \ 
\begin{enumerate}
\item Generate $ f ( \theta  |  \vec{\pi}_L  ; \vec{a} )$ from the FPT model with $\phi_{l,j} \sim Beta(1,1)$, for $l = 1, \cdots, L$ and  $j = 1, \cdots, 2^{l-1}$.
\item For $k = 1, \cdots, K$, generate  $\theta_k  \sim f ( \theta  |  \vec{\pi}_L  ; \vec{a} )$ .
\item For $k = 1, \cdots, K$, generate $X_k   \sim Binomial (m_k, P_k)$, with  $P_k = \oplogit^{-1} (\theta_k)$.
 \end{enumerate}
 \end{dfn}

Ferguson (1974) had already noted the conjugacy of the FPT model, for which
the conditional distribution of the step function density given $\vec{\theta}$ is FPT with updated hyper-parameter values,
 \begin{equation} \label{hBeta:pst}
\phi_{l, j} | \vec{\theta} \sim Beta( 1 + N_{l, 2 \cdot j - 1}, 1 + N_{l, 2 \cdot j}),
\end{equation}
for counts variables,  $N_{l, j} = | \{ k : \theta_k \in {\cal I}_{l, j} \} |$.
Let $f (\vec{\pi}_L | \vec{\theta})$ denote the density of the subinterval probabilities in the FPT model in (\ref{hBeta:pst}).
Expressing,
 \[
 f ( \vec{\pi}_L | \vec{X} )   
  = \int_{\vec{\theta}} f ( \vec{\pi}_L, \vec{\theta}  |  \vec{X})   d\vec{\theta}  
  = \int_{\vec{\theta}} f ( \vec{\pi}_L | \vec{\theta} ,  \vec{X})  f ( \vec{\theta}  |  \vec{X})   d\vec{\theta}  
  = \int_{\vec{\theta}} f ( \vec{\pi}_L | \vec{\theta})  f ( \vec{\theta}  |  \vec{X})   d\vec{\theta} , 
  \]
reveals that the conditional distribution of the step function density given $\vec{X}$ is a mixture of FPT models.
We evaluate this FPT mixture by the Gibbs sampler in Algorithm \ref{Gibbs-algo}.
To this end, we further derive the conditional distribution of  $\vec{\theta}$ given  $\vec{X}$ and $\vec{\pi}_L$,
\begin{eqnarray}
  f ( \vec{\theta}  |   \vec{X}  , \vec{\pi}_L)  & =  &    \frac{ f ( \vec{\theta},   \vec{X}   | \vec{\pi}_L)  \cdot f( \vec{\pi}_L)  }    { f( \vec{X}, \vec{\pi}_L)}    
   =  \frac{   \Pi_{k = 1}^K f ( \theta_k,  X_k   | \vec{\pi}_L)  \cdot f( \vec{\pi}_L)  }    { f( \vec{X}, \vec{\pi}_L) }    \label{cond.prb1} \\
& = &     \Pi_{k = 1}^K \left( f ( X_k   |  \theta_k  )   f ( \theta_k   | \vec{\pi}_L)  \right) \cdot \left(    f( \vec{\pi}_L)   /  f( \vec{X}, \vec{\pi}_L) \right). \label{cond.prb2}
 \end{eqnarray}
The second equality in (\ref{cond.prb1}) is because $(\theta_1, X_1), \cdots, (\theta_K, X_K)$ are conditionally independent given $\vec{\pi}_L$.
The equality in (\ref{cond.prb2}) is because given $\theta_k$, the distribution of $X_k$ does not depend on $ \vec{\pi}_L$.
Expression  (\ref{cond.prb2}) reveals that the components of $\vec{\theta}$ are conditionally independent 
with marginal conditional  density proportional to the product of the binomial likelihood of $X_k$ and the step function density that is determined by $ \vec{\pi}_L$.

\begin{algorithm}[]
	\SetAlgoLined
	\SetKwInOut{Set}{Set}
	\SetKwInOut{Input}{Input}
	\SetKwInOut{Output}{Output}
	
	Set number of Gibbs sampler iterations $G$,  initial values $\vec{\pi}^{(0)}_L$ 
	
	\vspace{.15cm}
	
	\For{$g = 1, \cdots,  G$}{
				
		\For{$k = 1,  \cdots, K$}{
			Sample  $\theta^{(g)}_k $ from its conditional  distribution given $\vec{X}$ and $\vec{\pi}^{(g-1)}_L$ in (\ref{cond.prb2}) }
		
		Sample $\vec{\pi}_L^{(g)}$ from its conditional distribution given $\vec{\theta}^{(g)}$ specified by (\ref{hBeta:pst})
	}

	\vspace{.15cm}
	
	Output posterior samples $\left( \vec{\theta}^{(1)}, \vec{\pi}^{(1)}_L \right),  \cdots,  \left( \vec{\theta}^{(G)}, \vec{\pi}^{(G)}_L \right)$

	\caption{Gibbs sampler for posterior distribution of Generative Model \ref{def:seq}}
	\label{Gibbs-algo}
\end{algorithm}

\medskip
\subsection{The deconvolution-based test}\label{subsec:deconv_based_tests}

For the deconvolution-based tests we consider the posterior distribution of the CDF of the mixing distribution $f ( \theta | \vec{\pi}_L ; \vec{a})$
at the endpoints of the dyadic partitions,  for which the CDF is given by the cumulative sum,
$CDF_{\vec{\pi}_L} (a_j) = \pi_{L,1} + \cdots + \pi_{L,j}$, for $j = 1, \cdots, 2^L$.
To incorporate the shift parameter $\rho \ge 0$ in the logit scale, 
we define ${\theta}^*_j (a_j, \rho) = a_{j} + \rho$,
which corresponds setting the success probability at which the null hypothesis is tested to $p^*_j =  \oplogit^{-1} ({\theta}^*_j)$.
Thus, we use the posterior median of $CDF_{\vec{\pi}_L} (a_{j})$, 
\begin{equation} \label{def-stat}
\widehat{CDF} (a_{j} ; \vec{X}) = \opmedian_{\vec{\pi}_L | \vec{X}}  \;  \left(  CDF_{\vec{\pi}_L} (a_{j}) \right),
\end{equation}
for testing the null hypothesis $H^{GE}_0 (q_0 ; p^*_j)$.
The significance level of the observed test statistic value is evaluated by Monte Carlo simulation  
of the test statistic distribution under the worst-case mixing distribution corresponding to $H^{GE}_0 (q_0 ; p^*_j)$,
 \begin{equation} \label{dfn-pval-LT-a}
\oppvalue  ( a_j , \rho, \vec{x} )  = Pr_{\vec{X} \sim \tilde{\pi}^{min} (p ; \ q_0, p^*_j ) }  \left(   \ \widehat{CDF}( a_{j} ; \vec{x}) \  \le  \ \widehat{CDF}( a_{j} ; \vec{X} )  \ \right).
\end{equation}

\medskip
\begin{prop} \label{prop66}
The test, 
reject   $H^{GE}_0 (q_0 ; p^*_j) $ if the p-value in (\ref{dfn-pval-LT-a}) is less than or equal to $\alpha$, has significance level $\alpha$.
\end{prop} 

\begin{pf}
Assume null hypothesis $H^{GE}_0 (q_0 ; p^*_j) $ is true.
Then per construction, $\tilde{\pi}^{min} (p ; q_0, p^*_j)$ is stochastically smaller than $\tilde{\pi} (p)$, the null mixing distribution that generated the data.
According  to Proposition \ref{prop11}  (statement and proof deferred to Appendix A), for all $t \in [0,1]$,
\begin{equation} \label{cor-exp}
 Pr_{\vec{X} \sim \tilde{\pi}^{min} (p ; q_0, p^*_j ) }  \left(  t \le \widehat{CDF}( a_{j} ; \vec{X} )  \right)
\ge  Pr_{\vec{X} \sim \tilde{\pi} (p) }  \left( t  \le \widehat{CDF}( a_{j} ; \vec{X})  \right).
\end{equation}
To complete the proof, setting $t = \widehat{CDF}( a_{j} ; \vec{x})$ in Expression (\ref{cor-exp})
reveals that the significance level for $\widehat{CDF}( a_{j} ; \vec{x})$ is less than or equal to  
$\oppvalue  ( a_j , \rho, \vec{x} )$  in (\ref{dfn-pval-LT-a}),  which is less than or equal to $\alpha$
if the test rejects the null hypothesis.
\end{pf}


\begin{exa}   \label{exa2}
{\bf Implementation of deconvolution-based tests}

\noindent
We illustrate the use of deconvolution-based tests for constructing left-tailed $95\%$ confidence intervals for the $q = 0.40$ quantile of the $Beta(2,2)$ mixing distribution, 
for the sample of binomial counts we considered in  Example \ref{exa1}.
To compute the deconvolution-based statistics we run $G = 1000$ iterations of Algorithm \ref{Gibbs-algo},
for a $L = 8$ level FPT model on endpoints vector $\vec{a}$,  with $a_j =  j \cdot 10 /  256   - 5$   for $j = 0, 1, \cdots, 256$.
The Gibbs sampling algorithm produces probability vectors, $\vec{\pi}^{(g)}_8  = ( \pi^{(g)}_{8, 1},  \cdots,  \pi^{(g)}_{8, 256})$,  
for $g = 1,  \cdots,  G$. Thus $\pi^{(g)}_{8, 1} + \cdots +\pi^{(g)}_{8, j}$ are realizations of the posterior distribution of $CDF_{\vec{\pi}_L} (a_j)$.

In the left panel of Figure 3 we display the results of the Gibbs sampling algorithm for the data considered in  Example  \ref{exa1}.
The blue and red curves are identical to the blue and red curves in Figure 2.
The black circles  and green curves display the $0.025$, $0.50$, $0.975$ quantiles of the posterior distribution of 
$CDF_{\vec{\pi}_L} (a_j)$, for $j = 0, \cdots, 256$, produced by the Gibbs sampler.
In the right panel of Figure 3 we display the results of the Gibbs sampling algorithm for a single sample, 
$\vec{X} \sim  \tilde{\pi}^{min} (p ;  0.40, 0.668)$, in which $P_k = 0$ for $34$ observations and $P_k = 0.668$ for the remaining $46$ observations.
The two plots display that the hierarchical Bayes approach yields tighter estimates for the mixing distribution than the empirical CDF of $\hat{P}_k$.
Thus even though in the right plot the deconvolution-based estimate for CDF of the mixing distribution at $0.668$ is approximately $0.70$,
the deconvolution-based estimate decreases to $0.425 = 34 / 80$ at larger success probabilities than the empirical CDF of $\hat{P}_k$,
allowing us to work with smaller shift parameter values for testing $H^{GE}_0 ( 0.40 ; 0.668 )$.
 
 As $a_{145} = 0.6641$ is the largest $a_j$ that is smaller than  $\oplogit(0.668) =  0.6991$,
 the smallest positive shift parameter value we may use for testing $H^{GE}_0 ( 0.40 ; 0.668 )$ is $\rho = 0.6991 - 0.6641 = 0.035$,
 with observed test statistic value, $\widehat{CDF} (a_{145} ; \vec{x})  = 0.753$.
To evaluate significance levels for testing $H^{GE}_0 ( 0.40 ; 0.668 )$,
we simulated $1000$ worst-case null mixture distribution samples, $\vec{X} \sim  \tilde{\pi}^{min} (p ;  0.40, 0.668)$.
For $99$ null samples $\widehat{CDF} (a_{145}  ; \vec{X})$ exceeded $0.753$,
yielding $\oppvalue  ( a_{145} , 0.035, \vec{x} )  = 0.099$.
To test $H^{GE}_0 ( q_0 = 0.40 ; p_0 = 0.668 )$ with $\rho \approx 0.50$, 
we evaluate the CDF at $a_{133} =  0.1953$ for which $\rho =  0.6991 - 0.1953  = 0.5038$.
For the sample of binomial counts displayed in the left panel of Figure 3,
 $\widehat{CDF} (a_{133} ; \vec{x}) = 0.578$.
In $2$ out of $1000$ null samples  $\widehat{CDF} (a_{133}  ; \vec{X})$ exceeded  $0.578$, 
yielding $\oppvalue  ( a_{133} , 0.5038, \vec{x} )  = 0.002$.
 
 The results of the previous paragraph suggest that the $95\%$ deconvolution-based left-tailed confidence intervals for the $q = 0.40$ quantile 
 is larger than $[0, 0.668]$ for $\rho = 0$ and smaller than $[0, 0.668]$ for $\rho = 0.50$.
 We use the `mcleod.estimate.CI.single.q' function in the {\em mcleod} package to construct these confidence intervals.
And indeed, for $\rho = 0$ the confidence interval was $[0,0.698]$,
for $\rho = 0.50$ the confidence interval was $[0, 0.659]$.
While setting $\rho = 0.10$ yielded the confidence interval $[0, 0.632]$.
 \end{exa}


\begin{figure}[] 
\centering
\includegraphics[height=0.95 \textwidth]{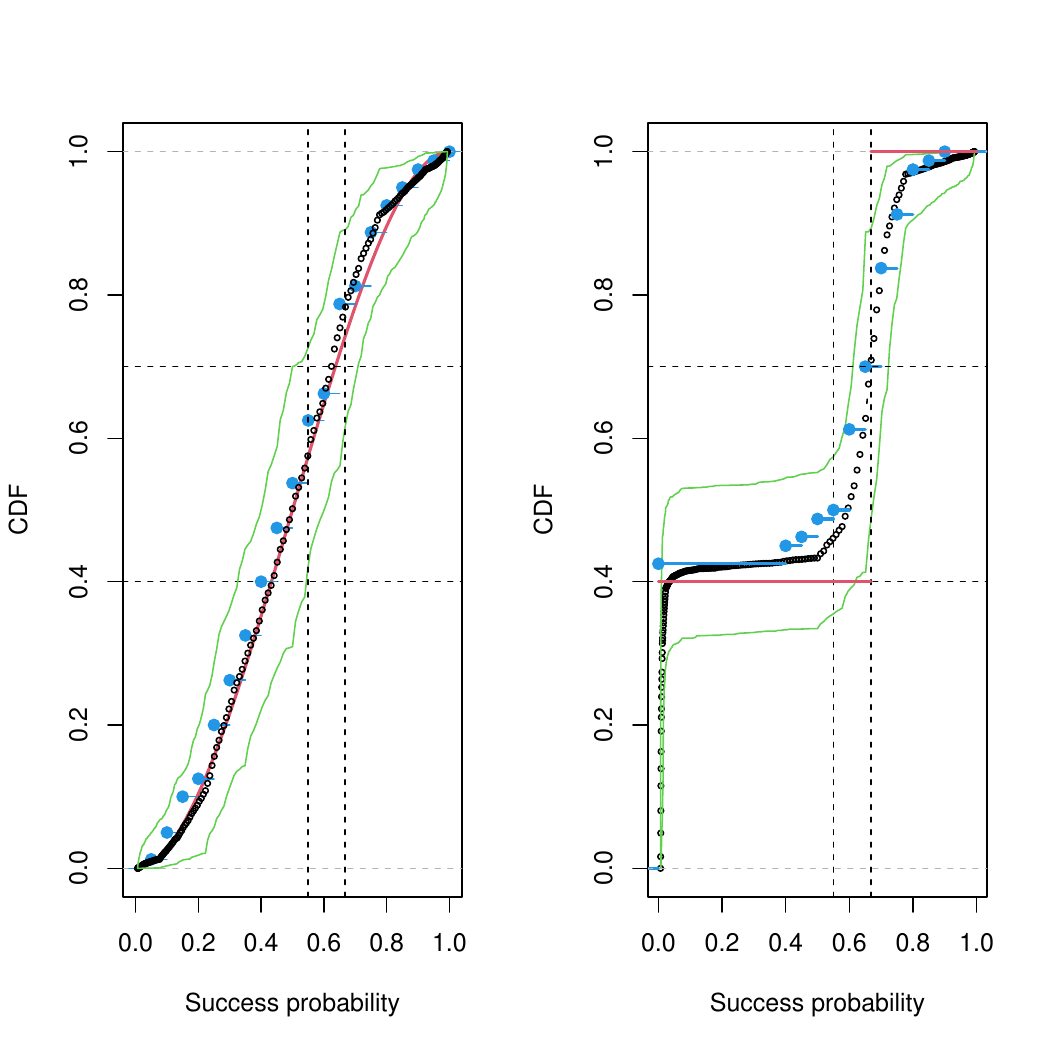}
\caption{
Simulated example -- deconvolution-based analysis for constructing  a 95\% confidence interval for the $0.40$ quantile of the mixing distribution.
In the left panel we display analysis of the $Beta(2,2)$ mixing distribution data we considered in Example \ref{exa1}.
In the right we display analysis for a realisation of  the $\tilde{\pi}^{min}(p; 0.40, 0.668)$ mixing distribution data.
The red curves are the CDF of the mixing distributions.
The blue  curves are the empirical CDF of $\hat{P}_1, \cdots, \hat{P}_{80}$.
The black points display the posterior median of the CDF of the mixing  
distribution at success  probability values $\oplogit^{-1}(a_j)$, for  $j = 0,  \cdots,  256$. 
The green curves display the $0.025$ and $0.975$ quantiles of the posterior distribution of the CDF of the mixing distribution.
The vertical lines are drawn at success probabilities $0.668$ and $0.55$.
The horizontal lines are drawn at CDF values $0, 0.40, 0.70, 1$.
}
\end{figure}


\bigskip
\subsection{Determining the value of the shift parameter} \label{det:rho}
The value of $\rho$ that yields the shortest confidence interval depends 
on the quantile for which the confidence interval is constructed, the shape of the mixing distribution,
 the number of binomial counts, and mainly on the sample sizes of the binomial counts.
 Algorithm \ref{algo3} is a data driven procedure that uses a subset of the observations for determining the value of $\rho$ and
the remaining observations for constructing a confidence interval for the $q_0$ quantile of the mixing distribution.

\begin{algorithm}[]
	\SetAlgoLined
	\SetKwInOut{Specify}{Specify}
	\SetKwInOut{Output}{Output}
		
	\vspace{.15cm}
	
	 Randomly select a subset, $S_1 \subset \{ 1 \cdots  K \}$,  comprising of 10\%-20\% of the samples
	and partition the data into calibration data $\vec{x}^1 = \{ x_k  : k \in S_1 \}$ and test data  $\vec{x}^2 = \{ x_k  : k \notin S_1 \}$.
	
	\vspace{.15cm}
	Use calibration data to compute test statistics values, $\widehat{CDF} (a_j ; \vec{x}^1)$ for $j = 1, \cdots, 2^L$.

	\vspace{.15cm}
	 For a sequence of shift parameter values, $\rho_1, \cdots, \rho_R$, construct confidence intervals for the $q_0$ quantile of $\tilde{\pi} (p)$ 
	on the basis of the following significance assessments,
	\[
		\oppvalue  ( a_j,  \rho_r, \vec{x}_1 )  = Pr_{\vec{X}^2 \sim \tilde{\pi}^{min} (p ; \ q_0, {\theta}^*_j (\rho_r) ) } 
		 \left(   \ \widehat{CDF}( a_{j} ; \vec{x}^1) \  \le  \ \widehat{CDF}( a_{j} ; \vec{X}^2 )  \ 	\right)
	\]
	where ${\theta}^*_j (\rho_r)  = a_j + \rho_r$ for $r = 1, \cdots, R$. Let $\rho_{min}$ the shift parameter value yielding the shortest confidence interval.

	\vspace{.15cm}
	Use test data to compute test statistics values, $\widehat{CDF} (a_j ; \vec{x}^2)$ for $j = 1, \cdots, 2^L$.

	\vspace{.15cm}
	 Output the confidence interval for the $q_0$ quantile of $\tilde{\pi} (p)$  based on,
	\[
		\oppvalue  ( a_j,  \rho_{min},\vec{x}^2 )  = Pr_{\vec{X}^2 \sim \tilde{\pi}^{min} (p ; \ q_0, {\theta}^*_{min} ) } 
				 \left(   \ \widehat{CDF}( a_{j} ; \vec{x}_2) \  \le  \ \widehat{CDF}( a_{j} ; \vec{X}^2 )  \ \right)
	\]
	for ${\theta}^*_{min}  = a_j + \rho_{min}$.

	\caption{Algorithm for constructing confidence intervals with optimal shift parameter value}
	\label{algo3}
\end{algorithm}

Expression $\vec{X}^2 \sim \tilde{\pi}^{min} (p ; \ q_0, {\theta}^* )$ in Algorithm \ref{algo3},
means that $\vec{X}^2 = \{ X_k : k \notin S_1 \}$ is generated by sampling  $X_k \sim Binomial ( m_k, P_k)$ with $P_k \sim \tilde{\pi}^{min} (p ; \ q_0, {\theta}^* )$.
Using this null sample in Step 5 is needed for the validity of the confidence intervals based on the test data.
We use  this null sample in Step 3 so that value of $\rho$ we derive, on the basis of the calibration set estimate of the mixing distribution, 
will yield a small test set confidence interval.

In Appendix \ref{const-curves} we suggest changes in Algorithm \ref{algo3} for selecting the value of $\rho$ for constructing confidence curves for the CDF.
To construct the confidence curves for  Example \ref{exa0}, shown in Figure 1,  we used $\rho = 0.16$.
In  Appendix \ref{subsec:appendix_Additional_Efron} we discuss selecting the shift parameter value in Example \ref{exa0},
and show that using part of the data for calibration has little effect on the tightness of  the confidence curves.


%
%
%
%


\bigskip

\bigskip
\section{Asymptotic behaviour of the confidence intervals}\label{sec:asymptotic_behaviour}

In this section we discuss the behaviour of the confidence intervals for the mixing distribution based on mixture distribution samples 
when $K$ tends to infinity.
In the case that $m_k \rightarrow \infty $ for all $k$, $\hat{P}_k = X_k / m_k$ converges to $P_k$,
thereby reducing the problem to a standard distribution estimation problem.
For which for $K \rightarrow \infty$, the difference of the CDF estimate based on the observed $\vec{X}$ and that of the estimate from the no-noise case vanishes at all continuity points of $\tilde{\pi} (\cdot)$.

\medskip
The more interesting regime is that $K \rightarrow \infty$ with $m_k \le m_{max}$ for all $k$.
We first show that the lack of identifiability of the mixing distribution by the mixture distribution,
limits the tightness of confidence intervals for quantiles of the mixing distribution.
Expressing the mixture distribution of $X_k \sim \tilde{\pi} (p)$, 
\begin{eqnarray*}
\Pr( X_k = j)  & = &  E_{P_k \sim \tilde{\pi}(p)} \left( {m_k \choose j} (1-P_k)^{m_k - j}  P_k^j \right) =
 E_{P_k \sim \tilde{\pi}(p)} \left(  {m_k \choose j} \left(\sum_{i=0}^{m_k -j} {m_k - j  \choose i}   (-P_k)^{i} \right)P_k^{j} \right)  \nonumber \\
& = &
{m_k \choose j} \sum_{i=0}^{m_k -j} {m_k - j  \choose i}  (-1)^{i} E_{P_k \sim \tilde{\pi}(p)} \left( P_k^{i+j} \right), 
\end{eqnarray*}
reveals that mixing distributions $\tilde{\pi}'$ and $\tilde{\pi}''$ with the same first $m_{max}$ moments  yield the same mixture distribution for all $X_k$.
Suppose $\tilde{\pi}'$ and $\tilde{\pi}''$ are two mixing distribution that produce the same mixture distribution for all $X_k$, 
with corresponding $q_0$ quantiles $p'$ and $p''$, such that $p' < p''$. 
As each mixture distribution sample, $X_k \sim \tilde{\pi}' (p)$ for $k = 1,  \cdots,  K$, also constitutes as a mixture distribution sample for 
mixing distribution $\tilde{\pi}'' (p)$,
then any $1 - \alpha$ left-tailed confidence interval for the $q_0$ quantile of $\tilde{\pi}'$ based on $\vec{X}$ will also cover $p''$
with probability $1 - \alpha$. 

\bigskip
\subsection{Smallest asymptotic confidence interval}

Leveraging the stochastic monotonicity property needed from the test statistics, 
we explicitly derive the smallest confidence interval that may be constructed for quantiles of a given mixing distribution.
For simplicity, in this subsection we assume that all binomial samples have the same sample size, i.e. $m_k = m$,  $\forall k = 1, \cdots, K$.
Let $p'$ be a $q_0$ quantile of $\tilde{\pi}' (p)$.
By construction, $\tilde{\pi} ^{min} (p ; q_0,  p')$ is stochastically smaller than $\tilde{\pi}' (p)$.
Therefore acording to Lemma \ref{lemm3},   $X_k \sim \tilde{\pi} ^{min} (p ; q_0,  p')$ is stochastically smaller than  $X_k \sim \tilde{\pi}' (p)$.
Let  $p_{max} = p_{max}(m)$   denote the maximal $p'' \in [0,1]$ for which  $X_k \sim \pi ^{min} (p ; q_0,  p'')$ 
is stochastically smaller than $X_k \sim \tilde{\pi}' (p)$.

\begin{prop} \label{prop2}
For any $K$, any left-tailed $1 - \alpha$ confidence interval for the $q_0$ quantile for mixing distribution $\tilde{\pi}' (p)$ based on $\vec{X}$,
presented in this manuscript, will cover  $[0, p_{max}]$ with probability greater than or equal to $1 - \alpha$.
\end{prop}

\begin{pf}
Let $1 > p_1 > p_2 > \cdots > p_N > 0$ denote the sequence of success probabilities used for constructing 
the $1-\alpha$ left-tailed confidence interval for the $q_0$ quantile of $\tilde{\pi}' (p)$.
If $P_N > p_{max}$ then the confidence interval covers   $[0, p_{max}]$ with probability $1$.
Otherwise let $p_{n'}$ denote the largest $p_n$ that is less than or equal to  $p_{max}$.
By construction, 
$X_k \sim \pi ^{min} (p ; q_0,  p_{n'})$ is stochastically smaller than $X_k \sim \tilde{\pi}' (p)$.
As the components of $\vec{X}$ are independent identically distributed, then $\vec{X} \sim \pi ^{min} (p ; q_0,  p_{n'})$ is stochastically smaller than $\vec{X} \sim \tilde{\pi}' (p)$.
Which implies that for the two types of test statistics employed in this paper,  
for any $K$ and any shift parameter value $0 < \rho$, 
the test statistic distribution for $\vec{X} \sim \tilde{\pi}^{min} (p ; q_0, p_{n'})$ is stochastically greater than the test statistic distribution for $\vec{X} \sim \tilde{\pi}' (p)$.
For the mixture distribution empirical CDF test statistic in (\ref{count2}) this holds per defintion;
for the deconvolution-based statistic (\ref{def-stat}) this property is a corollary of Lemmas \ref{lemm5b} and \ref{lemm4}.
To complete the proof, note that if $\tilde{\pi}^{min} (p ; q_0, p_{n'})$ yields a larger test statistic distribution than $\tilde{\pi}' (p)$,
then per construction the level $\alpha$ test for $H_0^{GE} (q_0 , p_{n'})$ is accepted with probability of at least $1 - \alpha$,
in which case the resulting confidence interval covers  $[0, p_{max}]$.
\end{pf}

\medskip
To justify calling $[0, p_{max}]$ the smallest asymptotic confidence interval, we provide an algorithm that for any $0 < \epsilon$ and sufficiently
large $K$, yields confidence interval $[0, p_{max} + \epsilon]$ for the $q_0$ quantile of $\tilde{\pi}'(p)$ with arbitrarily high probability.
For $p_{max} + \epsilon < 1$,  $\exists x_{\epsilon} \in \{ 0, 1, \cdots, m \}$ 
such that the CDF of $X_k \sim \pi ^{min} (p ; q_0,  p_{max} + \epsilon)$ at $x_{\epsilon}$, which we denote $\gamma_0 (p^*)$
 for $p^*  = x_{\epsilon} / m$,
 is smaller than the CDF of  $X_k \sim \tilde{\pi}' (p)$ at $x_{\epsilon}$,  which we denote $\gamma (p^*)$.
Let $\overline{CDF} (p^* ; \vec{X})$ denote the counts statistic in (\ref{count2}) at $p^*$.
We use $\overline{CDF} (p^* ; \vec{X})$  for testing $H_0^{GE} (q_0 , p_{max} + \epsilon)$ at level $\alpha$.
If the null hypothesis is rejected then the confidence interval is  $[0, p_{max} + \epsilon]$,
otherwise the confidence interval is  $[0, 1]$.

First of all, as $p_{max} + \epsilon$ is larger than $p'$, the $q_0$ quantile of $\tilde{\pi}'(p)$, this is indeed a valid $1 - \alpha$ confidence interval.
Next, under $H_0^{GE} (q_0 , p_{max} + \epsilon)$, the distribution of $\overline{CDF} (p^* ; \vec{X})$ is $Binomial ( \gamma_0( p^*), K ) / K$
and the null hypothesis is rejected for large test statistic values.
While for $X_k \sim \tilde{\pi}' (p)$,  the distribution of $\overline{CDF} (p^* ; \vec{X})$ is $Binomial ( \gamma( p^*), K ) / K$.
Thus, as $ \gamma_0( p^*) <  \gamma ( p^*)$, then for sufficiently large $K$ the significance level $\alpha$
test rejects the null hypothesis with arbitrarily high probability.

\begin{exa} \label{exa3}
In Figure $4$ we illustrate how we derive the smallest asymptotic confidence interval for the $0.40$ quantile of the $Beta(2,2)$ mixing distribution for $m_k = 20$.
As the $0.40$ quantile of the $Beta(2,2)$ is $0.433$ then mixing distributions $\tilde{\pi}^{min} (p ; 0.40,  0.50)$ and $\tilde{\pi}^{min} (p ; 0.40,  0.70)$ 
are not stochastically smaller than the $Beta(2,2)$ distribution.
The right plot reveals that  $X_k \sim \tilde{\pi}^{min} (p ; 0.40,  0.50)$ is stochastically smaller than  $X_k \sim Beta(2,2)$,
however as the CDF of $X_k \sim \tilde{\pi}^{min} (p ; 0.40,  0.70)$ at  $x = 9, 10, \cdots, 14$ is smaller than the CDF of $X_k \sim Beta(2,2)$,
there is no stochastic ordering between $X_k \sim \tilde{\pi}^{min} (p ; 0.40,  0.70)$ and $X_k \sim Beta(2,2)$.

For this example, the largest $p''$ for which $X_k \sim \tilde{\pi}^{min} (p ; 0.40,  p'')$ is stochastically smaller than  $X_k \sim Beta(2,2)$
is  $p_{max} (20) = 0.607$.
For intuition, we evaluated $p_{max}$ for the $Beta(2,2)$ mixing distribution for other values of $m_k$: 
$p_{max} (2) = 0.707$,
$p_{max} (5)= 0.674$,
$p_{max}  (200) = 0.511$, 
$p_{max}  (1000) = 0.473$.
Recall that for the Example \ref{exa1} data, with $K = 80$
the smallest $95\%$ confidence interval  for the $0.40$ quantile of the $Beta(2,2)$ based on $\vec{X}$ was $[0, 0.632]$, 
while the corresponding  $95\%$ confidence interval based on $\vec{P}$ was $[0,0.47]$. 
Thus even for $K = 1000$ the confidence interval based on mixture distribution samples 
 would be larger than the confidence interval for $K = 80$ mixing distribution samples.
 
For comparison, we evaluate the smallest asymptotic confidence for left-tailed confidence intervals for the  $0.40$ quantile of mixing distribution the $\tilde{\pi} ^{max} (p ; 0.41, 0.433)$,
which assigns probability $0.41$ to the event $P = 0.433$ and probability $0.69$ to the event $P = 1$.
The $0.40$ quantile of the $\tilde{\pi}^{max} (p ; 0.41, 0.433)$ is also $0.433$, 
but as the the $0.41$ quantile of this distribution is $1$, the smallest asymptotic confidence intervals for the $0.40$ quantile are considerably larger.
For the $\tilde{\pi} ^{max} (p ; 0.41, 0.433)$ mixing distribution with $m_k = 2$, $p_{max} = 1$.
Implying that for any $K$ and $m_k = 2$, the $95\%$ left-tailed confidence interval for the $0.40$ quantile of the mixing distribution 
is  $[0, 1]$ with probability $0.95$.
While 
 $p_{max} (5)= 0.998$,
$p_{max} (20) = 0.880$,
$p_{max}  (1000) = 0.506$. 
\end{exa}



\begin{figure}[] 
\centering
\includegraphics[height=.5\textheight,width= 1\textwidth]{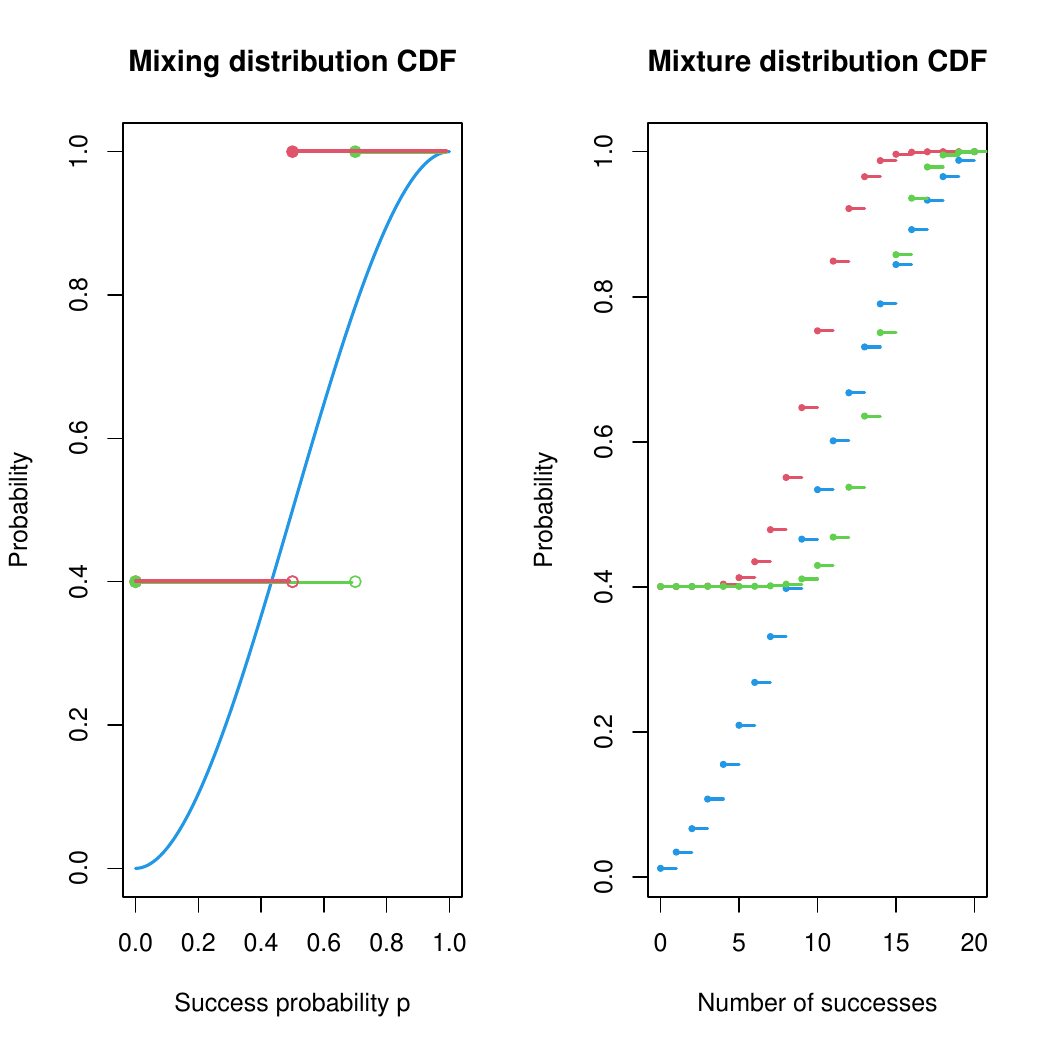}
\caption{Construction of smallest left-tailed confidence interval for the $0.40$ quantile of the $Beta (2,2)$ mixing distribution for $m_k \equiv 20$.
In the left panel we display the CDF of the mixing distributions.
In the right panel we display the CDF of $X_k \sim Binom(20,P_k)$,
The blue curves correspond to the $Beta(2,2)$ mixing distribution.
The red curves correspond to the $\tilde{\pi}^{min}(p; 0.40, 0.50)$ mixing distribution.
The green curves correspond to the $\tilde{\pi}^{min}(p; 0.40, 0.70)$ mixing distribution.
}
\end{figure}


\section{Discussion}
Our method for constructing confidence intervals for the mixing distribution 
may be applied to any estimation scheme that produces test statistics 
whose distribution stochastically increases if the mixing distribution is stochastically decreased.
We use finite Polya tree models because their high dimension and highly regularised hierarchical structure  
provides stable and accurate deconvolution estimates, for any type of mixing distributions, any configuration of binomial sample sizes, and any number of binomial counts, 
without the need for additional tuning parameters.
Interestingly, to derive the smallest asymptotic confidence interval we apply the counts statistic with an appropriately chosen shift parameter value.
In this work we only consider a binomial sampling distribution for $X_k$.
Our theoretical results apply to any mixture distribution that is likelihood-ratio increasing in $p_k$.
In the {\em mcleod}  R package we also allow $X_k$ to be Poisson.

\medskip
In the examples in this paper the components of the endpoints vector $\vec{a}$ are either a $65$ points, or a $257$ points, regular grid on $[-5,5]$.
Our numerical results suggest that, providing it is sufficiently dense, the choice of $\vec{a}$ has little effect on the resulting confidence intervals.
Computing the deconvolution-based estimates is relatively quick (less then a second for the data in Example \ref{exa0}).
To construct the confidence intervals we also need to specify the shift parameter value and compute the deconvolution-based estimates 
for multiple worst-case mixing distribution which may be considerably more time consuming:
the confidence curves for Example \ref{exa0} data required 13 minutes for computation on an i9-13900K PC.

\medskip
We have shown that incorporating a shift parameter is needed for constructing tight confidence interval for the mixing distribution
based on mixture distribution samples.
The data driven algorithms for determining the shift parameter values described in the text
are included in the {\em mcleod}  R package.

\medskip
The monotonicity property in Proposition \ref{prop11}, of the deconvolution-based test statistic (\ref{def-stat}) for Generative model 
\ref{def:seq} with $\phi_{l, j} \sim Beta (1,1)$,  
may be extended to the case that $\phi_{l, j} \sim Beta (\alpha_0, \alpha_0)$ with $\alpha_0 > 0$ at all hierarchy levels.
Our working experience suggests using $\alpha_0 = 1$.
However, this property does not hold for all Beta prior hyper-parameter values.
As a counterexample, we show that Lemma \ref{lemm5b}, which is a necessary condition for Proposition \ref{prop11},
does not hold for the FPT model with $L=2$,
 $\phi_{1,1} \sim Beta(10^6, 10^6)$, $\phi_{2,1} \sim Beta(1, 1)$, $\phi_{2,2} \sim Beta(1, 1)$ for $\vec{\theta}$ consisting of a single component $\theta_1$.
If  $\theta_1$ is in $ [a_1,a_2]$ then  $CDF_{\tilde{\pi}}(a_1)$ is a product of  $Beta(10^6 + 1, 10^6)$ and $Beta(1,2)$ random variables.
However, increasing $\vec{\theta}$, by moving  $\theta_1$ to $[a_2,a_3]$, changes $CDF_{\tilde{\pi}}(a_1)$ to the stochastically {\em larger}
product of $Beta(10^6, 10^6 + 1)$ and $Beta(1,1)$ random variables.


\medskip 
In Section 4, we derived the smallest asymptotic left-tailed confidence that may be
constructed by the tests presented in this paper, for the $q_0$ quantile of  a given mixture distribution for the binomial mixture distribution with equal sample sizes,
by showing that it is impossible to discriminate between the real mixing distribution and worst-case null mixing distributions corresponding to 
null hypotheses with larger $q_0$ quantiles that yield stochastically smaller mixture distributions.
As stochastic increase of the test statistic distribution by stochastic decrease of  the mixing distribution was necessary for constructing valid tests for $H_0^{GE} (q_0 , p)$.
This raises the question whether this result applies to any left-tailed confidence interval for the $q_0$ quantile of the mixing distribution constructed by inverting null
hypothesis $H_0^{GE} (q_0 , p)$.
And more generally, 
 without making parametric assumptions on the mixing distribution, 
 is it possible  to construct smaller left-tailed confidence intervals for quantiles of the mixing distribution 
from binomial mixture samples?

\begin{figure}[] 
\label{fig7}
\centering 
\includegraphics[height=.8\textheight,width=.8\textwidth]{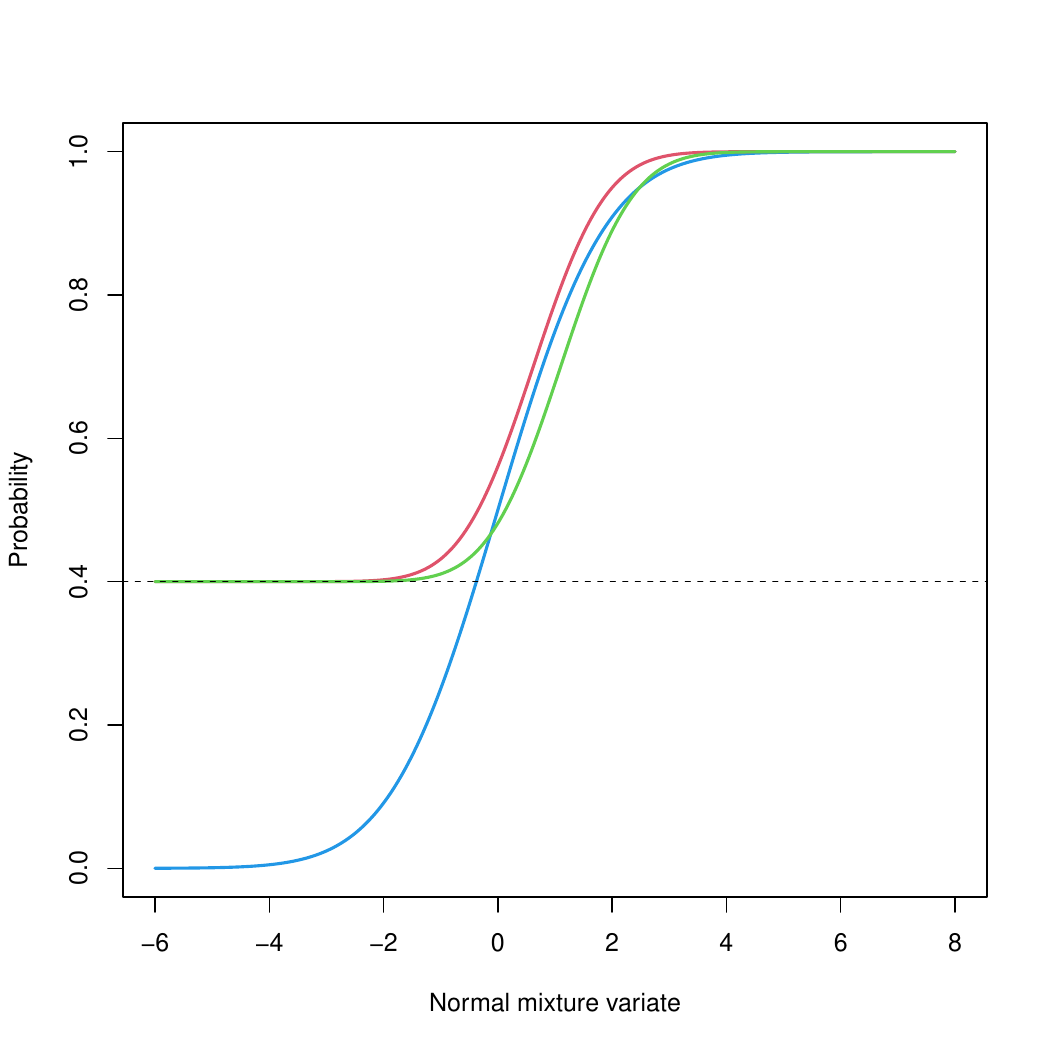}
\caption{CDF of $N ( \theta_k, 1)$ mixture distributions for $\theta_k = \oplogit( P_k)$.
The blue curve corresponds to $P_k \sim Beta(2,2)$.
The red curve and green curves correspond to worst case mixing distributions, $\tilde{\pi}^{min} (p ; 0.40, 0.65)$ and $\tilde{\pi}^{min} (p ; 0.40, 0.75)$.
}
\end{figure}

Note that it is also possible to derive smallest asymptotic confidence interval for the case that $K \rightarrow \infty$ and the binomial sample sizes, $m_k$, 
are sampled from a given finite distribution, and even for the case of continuous mixture distributions.
In Figure  \ref{fig7} we display the CDF for normal mixture distribution samples, $\hat{\theta}_k \sim N ( \theta_k, 1)$,
for $\theta_k =  \oplogit(P_k)$, with $P_k \sim Beta(2,2)$,  $P_k \sim  \tilde{\pi}^{min} (p ; 0.40, 0.65)$,  and $P_k \sim \tilde{\pi}^{min} (p ; 0.40,  0.75)$.
Note that similarly to the binomial mixture distributions shown in Figure 4,
there is no stochastic ordering between $\hat{\theta}_k \sim Beta(2,2)$ and  $\hat{\theta}_k \sim \tilde{\pi} ^{min} (p ; 0.40,  0.75)$,
however $\hat{\theta}_k  \sim \tilde{\pi}^{min} (p ; 0.40,  0.65)$ is stochastically smaller than $\hat{\theta}_k \sim Beta(2,2)$.
Implying that also for this case,
if a test statistic 
whose distribution stochastically increases if the mixing distribution stochastic decreases is used for constructing the confidence intervals,
then for any value of $K$ the left-tailed $0.95$ confidence interval for the $0.40$ quantile of the 
$Beta(2,2)$ distribution will cover $0.65$ with probability greater than $0.95$. 
And in general, it is also possible to specify the smallest asymptotic left-tailed confidence interval $[0, p_{max}]$
for the $q_0$ quantile of a given mixing distribution for normal mixture distribution samples.
For the $Beta (2,2)$ mixing distribution and $N ( \theta_k, 1)$ mixture distribution, shown shown in Figure \ref{fig7}, $p_{max} = 0.684$.
While for the $\tilde{\pi} ^{max} (p ; 0.41, 0.433)$ mixing distribution, considered in Example \ref{exa3}, and the $N ( \theta_k, 1)$ mixture distribution,
$p_{max} = 0.986$.

\appendix

\bigskip
\bigskip

\section{Monotonicity of the test statistic distribution}

In this  Appendix, ${\pi}(\theta)$, ${\pi}^1(\theta)$, ${\pi}^2 (\theta)$, denote mixing distributions
and  $\vec{X}$, $\vec{X}^1$, $\vec{X}^2$, denote the corresponding mixture distribution samples.
In this subsection we state three lemmas and then phrase and prove our monotonicity result for the deconvolution-based test statistic in (\ref{def-stat})
and Lemmas \ref{lemm4} and \ref{lemm3}, in the next subsection we prove Lemma \ref{lemm5b}.

\begin{lem} \label{lemm5b}
For $( \vec{\pi}_L, \vec{\theta},  \vec{X})$ generated in model \ref{def:seq},
increasing  $\vec{\theta}$ stochastically decreases the conditional distribution of $CDF_{\vec{\pi}_L}(a_j)$ given $\vec{\theta}$, for each $j   \in \{ 0, \cdots, 2^L \}$.
 \end{lem}

\begin{lem} \label{lemm4}
For $( \vec{\pi}_L, \vec{\theta},  \vec{X})$ generated in model \ref{def:seq},
the distribution of  $\vec{\theta} | \vec{X}  = \vec{x}$ stochastically increases in $\vec{x}$.
\end{lem}
 
\begin{lem} \label{lemm3} If ${\pi}^{1}(\theta)$ is stochastically greater than ${\pi}^{2}(\theta)$ then 
the distribution of $\vec{X}^{{1}}$  is stochastically greater than the distribution of $\vec{X}^{{2}}$.
\end{lem}

\begin{prop} \label{prop11}
For each $j \in \{ 0, \cdots, 2^L \}$,
if  ${\pi}^{1} (\theta)$ is stochastically greater than ${\pi}^{2} (\theta)$
then the distribution of  $\widehat{CDF}( a_{j} ; \vec{X}^1 )$  is stochastically smaller than the distribution of $\widehat{CDF}( a_{j} ; \vec{X}^2 )$.
\end{prop}

 
\begin{pf}{\it of  Proposition \ref{prop11}}.   Let $\vec{x}^1 = (x^1_1,  \cdots,  x^1_K)$ and 
$\vec{x}^2 = (x^2_1,  \cdots,  x^2_K)$ such that for all $k = 1,  \cdots,  K$, $x^1_k \ge x^2_k$.
Combining Lemmas \ref{lemm5b} and \ref{lemm4} yields that in model \ref{def:seq}, for each $j \in \{ 0,  \cdots, , 2^L \}$,
the conditional distribution of $CDF_{\vec{\pi}_L}(a_j)$  given $\vec{X} = \vec{x}^1$
is stochastically smaller than its conditional distribution given $\vec{X} = \vec{x}^2$.
Recall that
\[
\widehat{CDF} (a_{j} ; \vec{x}) = \opmedian_{\vec{\pi}_L | \vec{X} = \vec{x}}  \;  \{  CDF_{\vec{\pi}_L} (a_{j}) \}.
\]
This implies that $\widehat{CDF} (a_{j} ; \vec{x}^1) \le \widehat{CDF} (a_{j} ; \vec{x}^2)$.
And in general, $\widehat{CDF} (a_{j} ; \vec{x})$ is a decreasing function of $\vec{x}$.
Therefore according Lemma \ref{lemm3}, 
the test statistic distribution for mixing  distribution ${\pi}^{1}(\theta)$ is stochastically smaller than 
the test statistic distribution for mixing  distribution ${\pi}^{2}(\theta)$.

\end{pf}

\begin{pf}{\it of Lemma \ref{lemm4}}.
We begin by expressing the conditional distribution of  $\vec{\theta} | \vec{X}$,
\[
   f_{ \vec{\theta} | \vec{X}  =  \vec{x}} (  \vec{\theta})    = 
    \frac{   f_{ \vec{X}, \vec{\theta}} ( \vec{x},  \vec{\theta})}  {  f_{\vec{X}} ( \vec{x} )} = 
   \frac{   f_{  \vec{X} |  \vec{\theta}}  (   \vec{x} ) \cdot f_{   \vec{\theta}}  (    \vec{\theta})  } {  f_{\vec{X}} ( \vec{x} )}    = 
        \frac{   \Pi_{k = 1}^K  f_{X_k | \theta_k} ( x_k | \theta_k) \cdot  f_{   \vec{\theta}}  (    \vec{\theta}) }
         {   f_{\vec{X}} ( \vec{x})},
 \]
for  $f_{ \vec{\theta}}$ the marginal distribution of $\vec{\theta}$,
$f_{ \vec{X}}$ the marginal distribution of $\vec{X}$,
and  $f_{X_k | \theta_k}$  the binomial distribution of  $X_k | \theta_k$, which we express:
\[ 
f_{X_k | \theta_k} ( x_k | \theta_k)  =   {m_k \choose x_k}  \cdot  \left\{ \frac{ \exp(\theta_k)}{1+\exp(\theta_k)} \right\} ^{x_k} 
 \cdot \left\{1 -  \frac{ \exp(\theta_k)}{1+\exp(\theta_k)} \right\} ^{m_k-x_k}
\; = \;  {m_k \choose x_k}  \cdot  \frac{ \{ \exp(\theta_k) \}^{x_k} }{ \{ 1+\exp(\theta_k) \}^{m_k} }.
     \]
Therefore for $\vec{x}^1$ and $\vec{x}^2$, with $x^1_k \ge x^2_k$ $\forall k$, the ratio of the conditional densities of $\vec{\theta}$,
\begin{eqnarray*}
\frac{   f_{ \vec{\theta} | \vec{X}  = \vec{x}^1} (  \vec{\theta}) }{   f_{ \vec{\theta} | \vec{X}  = \vec{x}^2} (  \vec{\theta}) } & = & 
\frac{  \Pi_{k = 1}^K  f_{X_k | \theta_k} ( x^1_k | \theta_k) \cdot  f_{   \vec{\theta}}  (    \vec{\theta})  /   f_{\vec{X}} ( \vec{x}^1)}
{    \Pi_{k = 1}^K  f_{X_k | \theta_k} ( x^2_k | \theta_k) \cdot  f_{   \vec{\theta}}  (    \vec{\theta})  /   f_{\vec{X}} ( \vec{x}^2))}  \\
& = &       \frac{   \Pi_{k = 1}^K \{    {m_k \choose x^1_k}  \cdot  \exp(\theta_k)^{x^1_k} \} /   f_{\vec{X}} ( \vec{x}^1)}  
    {   \Pi_{k = 1}^K \{    {m_k \choose x^1_k}  \cdot  \exp(\theta_k)^{x^2_k} \} /   f_{\vec{X}} ( \vec{x}^2)  } \\
    & =  & \{  \Pi_{k = 1}^K    \exp(\theta_k)^{x^1_k - x^2_k} \}  \cdot  \frac{   \{ \Pi_{k = 1}^K    {m_k \choose x^1_k}  \} /   f_{\vec{X}} ( \vec{x}^1 )  } 
    {   \{ \Pi_{k = 1}^K    {m_k \choose x^2_k}  \} /    f_{\vec{X}} ( \vec{x}^2 )  },
       \end{eqnarray*}
is non-decreasing in each $\theta_k$,
 which implies that  $\vec{\theta} | \vec{X} = \vec{x}^1$ is stochastically greater than  $\vec{\theta} | \vec{X} = \vec{x}^2$.
 \end{pf}

\begin{pf}{\it of Lemma \ref{lemm3}}.
As the components of $\vec{X}^{1}$ and  $\vec{X}^{{2}}$ are independent it is sufficient to show that 
$\forall k$, $X^{{1}}_k$  is stochastically larger than $X^{{2}}_k$.
Note that $X^1_k | \theta^1_k = t$ and $X^2_k | \theta^2_k = t$ have the same binomial distribution.
For $t_2 < t_1$,  the ratio of the probability mass functions for this  distribution,
\[
\frac{ \Pr ( X_k = x |  \theta_k = t_1) }{ \Pr ( X_k = x |  \theta_k = t_2) } 
= \frac{ {m_k \choose x} \cdot (1 / (1+\exp(t_1))^{m_k} \cdot ( \exp(t_1))^{x }}
{{m_k \choose x} \cdot (1 / (1+\exp(t_2))^{m_k} \cdot ( \exp(t_2))^{x }} 
= \left\{  \frac{ 1+\exp(t_2)}
{  1+\exp(t_1)} \right\}^{m_k} \cdot  \exp(t_1 - t_2)^{x },
\]
is increasing in $x$. 
This implies that $X_k | \theta_k$ is stochastically increasing in $\theta_k$.
Thus $\forall b \in \mathbb{R}$, the CDF of $X_k | \theta_k$ at  $b$, $\Pr ( X_k \le b  | \theta_k)$,  
is a decreasing function of  $\theta_k$.
To complete the proof we express the CDF's of $X^1_k$ and $X^2_k$,
\[
\Pr ( X^1_k \le b )  = E_{\theta_k \sim \tilde{\pi}^1} \Pr ( X_k \le b  | \theta_k) \le E_{\theta_k \sim \tilde{\pi}^{2}} \Pr ( X^{2}_k \le b  | \theta_k) = \Pr ( X^{2}_k \le b ),
\]
where the inequality is because stochastically increasing the distribution  $\theta_k$  decreases the expectation of  decreasing functions in $\theta_k$.
\end{pf}


\bigskip
\subsection{Proof of Lemma \ref{lemm5b}}

\underline{The general idea} is that the conditional distribution of the probability vector $\vec{\pi}_L$ 
given $\vec{\theta}$ for the generative model \ref{def:seq} is determined by the indicator vector 
$\vec{\delta} = ( \delta_1, \cdots, \delta_K)$. Where $\delta_k$ is increasing in $\theta_k$, $\delta_k = | \{ j : a_j  < \theta_k \}|$. 
It is therefore sufficient to show that  for  $\vec{\delta}^1 = (\delta_1^1, \cdots, \delta_K^1)$  
and $\vec{\delta}^2 = (\delta_1^2, \cdots, \delta_K^2)$ such that $\forall k$  $\delta^1_k \le \delta^2_k$, $\forall j \geq 1$, 
the distribution of  $CDF_{\tilde{\pi}_L}(a_j) |\vec{\delta^2}$ is stochastically smaller than the distribution of  
$CDF_{\tilde{\pi}_L}(a_j) | \vec{\delta}^1$. As $\delta_k$ are integers, $\vec{\delta}^2$ can be derived from 
$\vec{\delta}^1$ by taking each component $\vec{\delta}^1_k$ and increasing it by $1$ until it is equal to $\delta^2_k$. 
Furthermore, as the distribution of $\vec{\pi}_L | \vec{\delta}$ is exchangeable in the ordering of the components of $\vec{\delta}$, for $j_0 = 1,  \cdots,2^L -1$,  
it is sufficient to consider $\vec{\delta}^1$ and $\vec{\delta}^2$, such that $\delta^1_1 = j_0$ and $\delta^2_1 = j_0 +1$ 
and $\forall k = 2 \cdots K$, $\delta^1_k =  \delta^2_k$. For our proof we express $CDF_{\tilde{\pi}_L}(a_j)  = \pi_{L,1} + \cdots + \pi_{L, j}$ 
and denote the common realized node counts for $\vec{\delta} = (\vec{\delta}^1, \vec{\delta}^2)$ by $N_{L, j} = n_{L, j}$.

\underline{For $L=1$} we have $\pi_{1,1} = \phi_{1,1}$, and as the configuration changes from 
$c_0 = (n_{1,1} + 1, n_{1,2})$ to $c_1 = (n_{1,1}, n_{1,2} + 1)$, the posterior distribution of $\pi_{1,1}$ 
changes from $\beta (n_{1,1} + 2, n_{1,2} + 1)$ to $\beta (n_{1,1} + 1, n_{1,2} + 2)$ which is stochastically decreasing. 
One simple way to see this is to couple the two conditional random variables via common independent gamma random variables 
$U = \Gamma(n_{1,1} + 1, 1)$, $V = \Gamma(n_{1,2} + 1, 1)$, and $W = \Gamma(1, 1)$. Then $\pi_{1,1}|c_0$ can be defined as 
$\pi_{1,1}|c_0 \eqd (U+W)/(U+V+W)$ and $\pi_{1,1}|c_1$ can be defined as $\pi_{1,1}|c_1 \eqd U/(U+V+W)$. As $W \geq 0$, 
in this definition it holds that pointwise  $(U+W)/(U+V+W) \geq U/(U+V+W)$. Thus $\pi_{1,1}|c_0 \stge \pi_{1,1}|c_1$. 
For the rest of this document we skip writing the rate parameter of the gamma variables as it is always understood to be 1. 
We follow a general `partial' coupling argument but we will not be able to pointwise couple the comparative random variables from $L \geq 2$.

\underline{For $L \geq 2$}, several of the many cases can be proved by induction. We assume that the stochastic ordering holds for level $L \leq l-1 $ and prove it for $L = l$. 
The leaf level has $2^l$ partitions. Suppose one observation moves from 
$j_0 < 2^{l-1}$ to $j_0 + 1$. Then note that the leading factor of the partial cumulative distributions for $(\pi_{l, 1}, \ldots,  \pi_{l, 2^{l-1}})$ which is $\phi_{1,1}$ 
do not change and is independent of the other terms. Therefore this change can be viewed 
as $L=l-1$ level tree and therefore stochastically decreasing by our assumption. Similarly for partial cumulative sums from $(\pi_{l, 2^{l-1}}, \ldots,  \pi_{l, 2^l})$, 
they can be written as $\phi_{1,1} + (1-\phi_{1,1}) \cdot \text {other independent terms}$, 
and therefore the same idea applies. When $j_0$ moves from $j_0 > 2^{l-1}$ to $j_0 + 1$, note that the partial cumulative distributions 
for $(\pi_{l, 1}, \ldots,  \pi_{l, 2^{l-1}})$ do not change. And the partial cumulative sums from $(\pi_{l, 2^{l-1}}, \ldots, 
\pi_{l, 2^l})$, they can be written as $\phi_{1,1} + (1-\phi_{1,1}) \cdot \text {other independent terms}$, for which $\phi_{1,1}$ do not change, 
and the changes in the independent terms can be viewed as a level $l-1$ tree and therefore stochastically 
decreasing. Therefore the only interesting case is when $j_0$ moves from $j_0 = 2^{l-1}$ to $j_0 + 1$.

Now let us look at the even-numbered partial sums, that is, any sum of even number of terms from the left at the leaf level of a $l$ level tree. 
Here the structure is the same that of the $l-1$ level tree when $j_0$ moves the $2^{l-2}$ node at level $l-1$ to the $2^{l-2} + 1$ node at level $l-1$. 
Therefore assuming that the stochastic ordering holds for level $l-1$, the even-numbered partial 
sums are stochastically decreasing. So we need to prove that the posterior distribution of the odd-numbered partial sums from the left in a level $l$ stochastically 
decreases when $j_0$ moves from $j_0 = 2^{l-1}$ to $j_0 + 1$. For example for $L=2$ 
we need to prove that when the configuration changes from $c_0 = (n_{2,1}, n_{2,2}+1, n_{2,3}, n_{2,4})$ to $c_1 = (n_{2,1}, n_{2,2}, n_{2,3}+1, n_{2,4})$, 
both $\pi_{2,1}$ and $\pi_{2,1} + \pi_{2,2} + \pi_{2,3}$ decreases. Note that $\pi_{2,1} + 
\pi_{2,2} + \pi_{2,3} = 1 - \pi_{2,4}$. Therefore by reasons of symmetry and reverse movement if we can show that $\pi_{2,1}$ decrease, 
then $\pi_{2,4}$ increases and therefore $\pi_{2,1} + \pi_{2,2} + \pi_{2,3}$ decreases. We note that the 
symmetry and reverse movement argument extends at all levels of the tree. Therefore we need to prove that the posterior distribution of the 
odd-numbered partial sums from the left until $\pi_{l, 2^{l-1}}$ in a level $l$ tree stochastically decreases 
when $j_0$ moves from $j_0 = 2^{l-1}$ to $j_0 + 1$. This will be the most complicated bit. Let us start with some easy examples.

\underline{For $L=2$} we want to show that $\pi_{2,1}$ decreases when the configuration changes from 
$c_0 = (n_{2,1}, n_{2,2}+1, n_{2,3}, n_{2,4})$ to $c_1 = (n_{2,1}, n_{2,2}, n_{2,3}+1, n_{2,4})$. We show this by partial coupling of 
$\pi_{2,1} = \phi_{1,1} \cdot \phi_{2,1}$. Express $\phi_{1,1}$ as before for level 2 where note $n_{1,1} = n_{2,1} + n_{2,2}$ and $n_{1,2} = n_{2,3} + n_{2,4}$. 
Define a new set of independent gamma random variables for $\phi_{2,1}$. Let $S = \Gamma(n_{2,1}  + 1)$, $R = \Gamma(n_{2,2} + 1)$, 
and $T = \Gamma(1)$. Then $\pi_{2,1}|c_0 \eqd [(U+W)/(U+W+V)][S/(S+R+T)]$ and $\pi_{2,1}|c_1 \eqd [U/(U+V+W)][S/(S+R)]$. 
Rewriting $\pi_{2,1}|c_0 \eqd [(U+W)/(U+W+V)][S/(S+R)][(S+R)/(S+R+T)]$ and $
\pi_{2,1}|c_1 \eqd [(U+W)/(U+V+W)][S/(S+R)][U/(U+W)]$. Note that we have expressed each of the terms as a product of three independent terms. 
The independence follows because $[S/(S+R)]$ is independent of $(S+R)$ and $[U/(U+W)]$ is  independent of $(U+W)$. 
And we have been able to couple the first two terms, that is they are the same for the two expressions. 
Therefore we only need to compare $[(S+R)/(S+R+T)] \sim \beta(n_{1,1} + 2, 1)$ with $[U/(U+W)] \sim \beta(n_{1,1}+1, 1)$ 
to determine the stochastic ordering which is decreasing. Note that these terms cannot be compared pointwise and hence we view the proof technique as `partial coupling'. 
Similarly for $L=3$ we want to show that $\pi_{3,1}$ and $\pi_{2,1} + 
\pi_{3,3}$ decreases when we change $j_0$ from leaf 4 to leaf 5 at level 3. For $\pi_{3, 1} = \pi_{2, 1} \cdot \phi_{3,1}$ and we know that $\phi_{3, 1}$ 
does not change and is independent of $\pi_{2, 1}$ which we have shown decreases. Thus $\pi_{3, 1}$ decreases. To show that $\pi_{2,1} + \pi_{3,3}$ 
decreases we need more work, elaborated in the several paragraphs below.

\underline{For $L=3$} we want to show that the posterior distribution of $\pi_{2,1} + \pi_{3,3}$ decreases when we change $j_0$ from 4 to 5. 
Let the configurations in the partitions change from $c_0 := (n_1, n_2, n_3, n_4+1, n_5, \ldots)$ to $c_2 :=  (n_1, n_2, n_3, n_4, n_5+1, \ldots)$. 
Define an intermediate configuration $c_1 := (n_1, n_2+1, n_3, n_4, n_5, \ldots)$.  We characterize the beta random variables through gamma random variables. 
The random variables of interest to us are $\pi_{2,1}|c_0$, $\pi_{2,1}|c_2$, $\pi_{3,3}|c_0$, $\pi_{3,3}|c_1$, and $\pi_{3,3}|c_2$. 
First we compare $\pi_{2,1}|c_0 + \pi_{3,3}|c_0$ with $\pi_{2,1}|c_0 + \pi_{3,3}|c_1$ and then we compare 
$\pi_{2,1}|c_0 + \pi_{3,3}|c_1$ with $\pi_{2,1}|c_2 + \pi_{3,3}|c_2$. 

Write $\pi_{2,1}|c_0 = \beta(\sum_1^4 n_j+2, \sum_5^8 n_j+1) \beta(n_1 + n_2 + 1, n_3 + n_4 + 2)$. Define 
$U = \Gamma(\sum_1^4 n_j+1)$, $W = \Gamma(1)$, and $V = \Gamma(\sum_5^8 n_j+1)$ independent. Similarly define $S = \Gamma(n_1 + n_2 + 1)$, 
$R = \Gamma(n_3 + n_4 + 1)$, $T = \Gamma(1)$ all independent. Then $\pi_{2,1}|c_0 = [(U+W)/(U+W+V)][S/(S+R+T)]$. 
Using the same set of gamma random variables but preserving the independence within the term $\pi_{2,1}|c_2 = [U/(U+V+W)][S/(S+R)]$.

Similarly write $\pi_{3,3}|c_0 = \beta(\sum_1^4 n_j+2, \sum_5^8 n_j+1) \beta(n_3 + n_4 + 2, n_1 + n_2 + 1) \beta(n_3+1, n_4+2)$. 
Define independent $M = \Gamma(n_3+1)$, $N = \Gamma(n_4+1)$, and $O = \Gamma(1)$. Thus $\pi_{3,3}|c_0 = [(U+W)/(U+W+V)][(R+T)/(R+T+S)][M/(M+N+O)]$. 
And $\pi_{3,3}|c_2 = [U/(U+W+V)][R/(R+S)][M/(M+N)]$, where $\pi_{3,3}|c_1 = [(U+W)/(U+W+V)][R/(R+S+T)][M/(M+N)]$. 

Note $\pi_{3,3}|c_0$ can be rewritten as $[(U+W)/(U+W+V)][(R+T)/(R+T+S)][M/(M+N)][(M+N)/(M+N+O)]$ and $\pi_{3,3}|c_1$ 
as $[(U+W)/(U+W+V)][(R+T)/(R+T+S)][M/(M+N)][R/(R+T)]$. Note that we have matched the product of four terms up to the first three terms and the products are independent. 
This follows from the fact that $R/(R+T)$ is independent of $R+T$, a property of beta and gamma random variables, similarly $M/(M+N)$ is independent of $M+N$. Now to compare $(M+N)/
(M+N+O)$ and $R/(R+T)$, the former is $\beta(n_3+n_4+2, 1)$ and the latter is $\beta(n_3+n_4+1, 1)$ and therefore the former is larger.

Note that $\pi_{2,1}|c_0$ can be rewritten as $[(U+W)/(U+W+V)][1 - (R+T)/(S+R+T)]$ and is therefore independent of the comparative 
final terms discussed in the previous sentence. Therefore $\pi_{2,1}|c_0 + \pi_{3,3}|c_0$ is stochastically larger than $\pi_{2,1}|c_0 + \pi_{3,3}|c_1$. 
To compare $\pi_{2,1}|c_0 + \pi_{3,3}|c_1$ with $\pi_{2,1}|c_2 + \pi_{3,3}|c_2$ we require several more rewrites but the idea is the same.

Rewrite $\pi_{2,1}|c_0 = [(U+W)/(U+W+V)][1 - R/(R+S)][(R+S)/(R+S+T)]$ and $\pi_{2,1}|c_2 = [(U+W)/(U+W+V)][1 - R/(R+S)][U/(U+W)]$. 
Similarly $\pi_{3,3}|c_1 = [M/(M+N)][(U+W)/(U+W+V)][R/(R+S)][(R+S)/(R+T+S)]$ and $\pi_{3,3}|c_2 = [M/(M+N)][(U+W)/(U+W+V)][R/(R+S)][U/(U+W)]$. 
Therefore $\pi_{2,1}|c_0 + \pi_{3,3}|c_1 = \{ [(U+W)/(U+W+V)][1 - R/(R+S)] + [M/(M+N)][(U+W)/(U+W+V)][R/(R+S)] \} [(R+S)/(R+T+S)]$ 
and $\pi_{2,1}|c_2 + \pi_{3,3}|c_2 = \{  [(U+W)/(U+W+V)][1 - R/(R+S)] + [M/(M+N)][(U+W)/(U+W+V)][R/(R+S)] \} [U/(U+W)]$. 
Note that in the two product of two terms the first term is the same and independent of the second terms. 
The second terms are $(R+S)/(R+T+S)$ and $U/(U+W)$ which are $\beta(\sum_1^4 n_j+2, 1)$ and $\beta(\sum_1^4 n_j+1, 1)$ respectively, and therefore the former is stochastically larger.

Now we return to our most complicated bit \underline{for the general $l$}, that is, we need to prove that the posterior distribution of the 
odd-numbered partial sums from the left until $\pi_{l, 2^{l-1}}$ in a level $l$ tree stochastically decreases when $j_0$ moves from $j_0 = 2^{l-1}$ to $j_0 + 1$. 
We need more of inductions of the more straightforward steps that we have observed, particularly for that of levels 2 and 3. 
First we start with a simple observation, note that for odd partial sums from left until $\pi_{l, 2^{l-2}}$, the partial sums may be written as 
$\pi_{2,1} \cdot \text{independent terms}$. Further note that $\pi_{2,1}$ decreases, as shown before and the independent terms do not change in this move.
 Therefore we do not have anything to prove. In spirit, this is why it was easy to show that $\pi_{3, 1}$ decreases for a level 3 tree. 
 For other partial sums, we potentially need a sequence of intermediate jumps, similar to what we needed to show that $\pi_{2,1} + \pi_{3,3}$ decreases for a level 3 tree.

For another half of the odd partial sums that is $\pi_{l, 2^{l-2}}$ through $\pi_{l, 2^{l-2}+2^{l-3}}$, they can be written as 
$\pi_{2, 1} + \pi_{3,3} \times \text{independent terms}$. For example, at level 4, it holds that $\pi_{4,1} + \cdots + \pi_{4,5} = \pi_{2,1} + \pi_{3,3} \cdot \phi_{4,5}$. 
Similarly, for level 5, it holds that $\pi_{5,1} + \cdots + \pi_{5,9} = \pi_{2,1} + \pi_{3,3} \cdot \phi_{4,5} \phi_{5, 9}$ and 
$\pi_{5,1} + \cdots + \pi_{5,11} = \pi_{2,1} + \pi_{3,3} \cdot \{ \phi_{4,5} + \phi_{4,6} \phi_{5, 11}\}$. For all these cases, we may approach the proof similar to that for level 3. 
That is define an intermediate step $c_1$ as the configuration moves from $c_0$ to $c_1$ to $c_2$. In this case the move is defined from $j_0 = 2^{l-1}$ to $2^{l-2}$ 
and then from $2^{l-2}$ to $2^{l-1} + 1$. Then we first compare $\pi_{2,1}|c_0 + \pi_{3,3}|c_0 \cdot Z$ 
with $\pi_{2,1}|c_0 + \pi_{3,3}|c_1 \cdot Z$ and further $\pi_{2,1}|c_0 + \pi_{3,3}|c_1 \cdot Z$ with $\pi_{2,1}|c_2 + \pi_{3,3}|c_2 \cdot Z$. Note that 
the independent random variable do not change with positions $c_0$, $c_1$, or $c_2$. Therefore $Z$ simply gets multiplied in the 
constructions of the random variables in the proof and does not alter the distribution of the comparative terms.

Suppose these ideas of several jumps hold until the tree of level $l-1$, then consider the partial sum 
$\pi_{l, 1} + \cdots \pi_{l, 2^{l-1}-3} = \pi_{2,1} + \pi_{3,3} + \ldots + \pi_{l-2, 2^{l-3}-1} + \pi_{l-1, 2^{l-2}-1} \cdot \phi_{l, 2^{l-1}-3}$. As an example, 
$\pi_{5,1} + \cdots + \pi_{5, 13} = \pi_{2,1} + \pi_{3,3} + \pi_{4,7} \cdot \phi_{5, 13}$. In this case several moves would be required namely, f
rom $j_0 = 2^{l-1}$ to $2^{l-1}-4$ to $2^{l-1}-8$ to $2^{l-1}-16$ to until $2^{l-2}$ and then from $2^{l-2}$ to  $2^{l-1} + 1$. 
For example from partition $(5, 16)$ to $(5, 12)$ to $(5, 8)$ to $(5, 17)$. For the change of $\pi_{5,1} + \cdots + \pi_{5, 13}$ 
this move has the same impact as that of the moves $(4, 8)$ to $(4, 6)$ to $(4, 4)$ to $(4, 9)$ has on $\pi_{2,1} + \pi_{3,3} + \pi_{4,7}$. 
This is because $\phi_{5, 13}$ is independent of all the other terms and its posterior does not change with the sequence of these moves. 
Therefore the only last bit that remains for us to show is that $\pi_{l, 1} + \cdots \pi_{l, 2^{l-1}-1}$ decreases when $j_0$ moves from $2^{l-1}$ to $2^{l-1} + 1$.

The partial sum $\pi_{l, 1} + \cdots \pi_{l, 2^{l-1}-1} = \pi_{2,1} + \pi_{3,3} + \ldots + \pi_{l-2, 2^{l-3}-1} + \pi_{l-1, 2^{l-2}-1} + \pi_{l-1, 2^{l-2}} \cdot \phi_{l, 2^{l-1}-1}$. 
In this case again we need several moves from $j_0 = 2^{l-1}$ to $2^{l-1}-2$ to  $2^{l-1}-4$ to $2^{l-1}-8$ to $2^{l-1}-16$ to until $2^{l-2}$ 
and then from $2^{l-2}$ to $2^{l-1} + 1$. We rewrite the sum as 
$\pi_{l, 1} + \cdots \pi_{l, 2^{l-1}-1} = \pi_{2,1} + \pi_{3,3} + \ldots + \pi_{l-2, 2^{l-3}-1} + \pi_{l-2, 2^{l-3}} \cdot \{ \phi_{l-1, 2^{l-2}-1} +  \phi_{l-1, 2^{l-2}} \cdot \phi_{l, 2^{l-1}-1} \}$. 
Note that in the first move, that is, $j_0 = 2^{l-1}$ to $2^{l-1}-2$ only the last term changes that is $\{ \phi_{l-1, 2^{l-2}-1} +  \phi_{l-1, 2^{l-2}} \cdot \phi_{l, 2^{l-1}-1} \}$ changes and all other 
terms does not change the distribution of this term. And we know from the first part of the proof of level 3, 
that $\{ \phi_{l-1, 2^{l-2}-1}|c_0 +\phi_{l-1, 2^{l-2}} \cdot \phi_{l, 2^{l-1}-1}|c_0 \}$ is stochastically larger than  $\{ \phi_{l-1, 2^{l-2}-1}|c_0 +  \phi_{l-1, 2^{l-2}} \cdot \phi_{l, 2^{l-1}-1}|c_1 \}$. 
So now we continue the move from $j_0 = 2^{l-1}-2$ to $2^{l-1}-4$ to $2^{l-1}-8$ to $2^{l-1}-16$ to until $2^{l-2}$ and then from $2^{l-2}$ to $2^{l-1} + 1$.

Let us evaluate where we are now, we are at $\pi_{2,1}|c_0 + \pi_{3,3}|c_0 + \ldots + \pi_{l-2, 2^{l-3}-1}|c_0 + \pi_{l-1, 2^{l-2}-1}|c_0 +  \pi_{l-1, 2^{l-2}} \cdot \phi_{l, 2^{l-1}-1}|c_1$. 
Note that any of the further moves does not change the distribution of $\phi_{l, 2^{l-1}-1}$ and it is independent of all of the other terms. That is we may view the partial sums as 
$\pi_{2,1}|c_1 + \pi_{3,3}|c_1 + \ldots + \pi_{l-2, 2^{l-3}-1}|c_1 + \pi_{l-1, 2^{l-2}-1}|c_0 +  \pi_{l-1, 2^{l-2}} \cdot \phi_{l, 2^{l-1}-1}|c_1$. Note that  we have changed the conditional to $c_1$ 
until level $l-2$ as for those levels it does not matter. The second move to $c_2 = 2^{l-1}-4$ is similar to the second part of the proof for level 3 but there are more complications. We want to compare the 
partial sums to 
$\pi_{2,1}|c_2 + \pi_{3,3}|c_2 + \ldots + \pi_{l-3, 2^{l-4}-1}|c_2 + \pi_{l-3, 2^{l-4}}|c_2 \cdot \{\phi_{l-2, 2^{l-3}-1}|c_1 + 
\phi_{l-2, 2^{l-3}}|c_2 \cdot \{ \phi_{l-1, 2^{l-2}-1}|c_2 + \phi_{l-1, 2^{l-2}} \cdot \phi_{l, 2^{l-1}-1}|c_2 \} \}$. We try to understand through an example.

Suppose we want to study the partial sum $\pi_{2,1} + \pi_{2,2} \cdot \{ \phi_{3,3} + \phi_{3,4} \cdot \phi_{4,7}\}$ for level 4, as we move from partitions 
$c_0 = \mathcal{P}_8$ to $c_1 = \mathcal{P}_6$ to $c_2 = \mathcal{P}_4$ to $c_3 = \mathcal{P}_9$. The move from $c_0$ to $c_1$ was already discussed. 
For the move from $c_1$ to $c_2$ we want to compare $\phi_{1,1}|c_2 \cdot \{ \phi_{2,1}|c_1 + \phi_{2,2}|c_1 \cdot \{ \phi_{3,3}|c_0 + \phi_{3,4} 
\cdot \phi_{4,7}|c_1\}\}$ to $\phi_{1,1}|c_2 \cdot \{ \phi_{2,1}|c_1 + \phi_{2,2}|c_2 \cdot \{ \phi_{3,3}|c_2 + \phi_{3,4} \cdot \phi_{4,7}|c_2\}\}$. 
In all of these terms $\phi_{1,1}$ do not change, and is independent so we can ignore for now. 
We stick to the same notations as before for the gamma  random variables.
$ \phi_{2,1}|c_1 \eqd S/(R+S+T)$, $\phi_{2,2}|c_1 \cdot \phi_{3,3}|c_0 \eqd [(R+T)/(R+S+T)] \cdot [M/(M+N+O)] = [(R+T)/(R+S+T)] \cdot [M/(M+N)] \cdot [(M+N)/(M+N+O)]$, 
and $\phi_{2,2}|c_2 \cdot \phi_{3,3}|c_2 \eqd [R/(R+S+T)] \cdot [M/(M+N)] = [(R+T)/(R+S+T)] \cdot [M/(M+N)] \cdot [R/(R+T)]$. 
Similarly $\phi_{2,2}|c_1 \cdot \{\phi_{3,4} \cdot \phi_{4,7}\}|c_1 \eqd [(R+T)/(R+S+T)] \cdot [N/(M+N+O)] \cdot \phi_{4,7}|c_1 = [(R+T)/(R+S+T)] \cdot [N/(M+N)] \cdot \phi_{4,7}|c_1 \cdot [(M+N)/(M+N+O)]$ 
and $\phi_{2,2}|c_2 \cdot \{\phi_{3,4} \cdot \phi_{4,7}\}|c_2 = [R/(R+S+T)][N/(M+N)] \cdot \phi_{4,7}|c_2 = [(R+T)/(R+S+T)][N/(M+N)] \cdot \phi_{4,7}|c_2 \cdot [R/(R+T)]$. 
Therefore $\phi_{2,1}|c_1 + \phi_{2,2}|c_1 \cdot \{ \phi_{3,3}|c_0 + \phi_{3,4} \cdot \phi_{4,7}|c_1\} \eqd [1-(R+T)/(R+S+T)] + \{ [(R+T)/(R+S+T)] \cdot [M/(M+N)] + [(R+T)/(R+S+T)] 
\cdot [N/(M+N)] \cdot \phi_{4,7}|c_1\} \cdot [(M+N)/(M+N+O)]$ 
and $\phi_{2,1}|c_1 + \phi_{2,2}|c_2 \cdot \{ \phi_{3,3}|c_2 + \phi_{3,4} \cdot \phi_{4,7}|c_2\} \eqd [1-(R+T)/(R+S+T)] +
 \{ [(R+T)/(R+S+T)] \cdot [M/(M+N)] + [(R+T)/(R+S+T)] \cdot [N/(M+N)] \cdot \phi_{4,7}|c_2\} \cdot [R/(R+T)]$. 
We note that $\phi_{4,7}$ does not care if we condition on $c_1$ or on $c_2$. The comparative terms here are $[(M+N)/(M+N+O)]$ and $[R/(R+T)]$. 
And we note that these comparative terms are independent of $[M/(M+N)]$, $[N/(M+N)]$, and $(R+T)$. 
As in step 1 of the proof of level 3 the former comparative term is stochastically larger.

For general $l$ repeated (but convoluted) application of these phenomena applies. 
We observe how the previous graph is different from the first part of the proof of level 3. The term $\{[1-M/(M+N)] \cdot \phi_{4,7}|c_2\}$ 
was added to $M/(M+N)$. So we observe that this would be the induction structure, that is compared to a level $l-1$, 
in the second from right passive term a term will be added which will have the structure of one minus that term multiplied by an 
independent non-changing random variable. And we had already known that the comparative terms are independent of this passive term, therefore this will be independent of this additional term too. 
This seems to be the idea, that is, first move from the $2^{l-1}$ to $2^{l-1} -2$, and show that this is stochastically decreasing using the first part of the proof of level 3. 
This enables that the last term of the last odd leaf is independent and invariant to 
future moves. Then use the improvisations of the already available expressions of the $l-1$ level tree. 


\bigskip
\bigskip

\section{Additional algorithms}

We provide algorithms for constructing one-sided confidence intervals for the CDF of $\tilde{\pi} (p)$,
right-tailed confidence intervals for quantiles of $\tilde{\pi}  (p)$,
two-sided confidence intervals for the CDF and quantiles of $\tilde{\pi}  (p)$.
We also provide an algorithm for performing deconvolution-based tests for $H^{LE}_0 ( q ; p) :  q \le CDF_{\tilde{\pi}} ( p)$ 
and theoretical justification for its validity.

\medskip
\subsection{Left-tailed confidence intervals for the CDF} 
To construct a left-tailed confidence interval for $CDF_{\tilde{\pi}} (p_0)$, we consider a sequence of decreasing  CDF values $1 > q_1 > q_2 > \dots > q_N > 0$
and consecutively test null hypotheses $H^{LE}_0 ( q_n ; p_0)$ at level $\alpha$. 
Beginning with $n = 1$, we proceed to test $H^{LE}_0 ( q_{n+1} ; p_0)$  if $H^{LE}_0 ( q_n ; p_0)$ is rejected. Testing is 
stopped once a null hypothesis is accepted or all the null hypotheses have been tested and rejected.
If $H^{LE}_0 ( q_1 ; p_0)$ is accepted, the confidence interval for $CDF_{\tilde{\pi}} (p_0)$ is  $[ 0, 1]$.
Otherwise,  the left-tailed CI for $CDF_{{\pi}} (p_0)$ is $[0, q_{ul}]$,  where $q_{ul}$ 
is the smallest $q_n$ for which $H^{LE}_0 ( q_n ; p_0)$  was rejected.
 
We show that this is a valid $1 - \alpha$ confidence interval for $CDF_{\tilde{\pi}} (p_0)$.
First, note that if $CDF_{\tilde{\pi}} (p_0) < q_N$ then $CDF_{\tilde{\pi}} (p_0)$ is covered by the confidence interval with probability $1$.
Otherwise, let $q_{crit}$ be the largest $q_n$ that is smaller than $CDF_{\tilde{\pi}} (p_0)$.
Note that  $CDF_{\tilde{\pi}} (p_0)$ is not covered by the confidence interval only if  $H^{LE}_0 (q_{crit}; p))$ is rejected.
As $ q_{crit} < CDF_{\tilde{\pi}} (p_0)$, $H^{LE}_0 (q_{crit}  ; p_0)$ is a true null hypotheses, thus the probability this occurs is $\le \alpha$.

\medskip
\subsection{Right-tailed confidence intervals for the CDF} 

For right-tailed confidence intervals for $CDF_{\tilde{\pi}} (p_0)$, we consider a sequence of increasing  CDF values $0 < q_1 <  q_2 < \dots < q_N < 1$.
For $n = 1, \cdots, N$, we consecutively test $H^{GE}_0 ( q_n ; p_0)$ at level $\alpha$, until the null hypothesis is accepted.
If $H^{GE}_0 ( q_1 ; p_0)$ is accepted, the confidence interval for $CDF_{\tilde{\pi}} (p_0)$ is  $[ 0, 1]$;
otherwise, it is $[q_{ll},1]$,  where $q_{ll}$ is the largest $q_n$ for which $H^{GE}_0 ( q_n ; p_0)$  was rejected.

\medskip
\subsection{Right-tailed confidence intervals for quantiles} 
For right-tailed confidence interval for the $q_0$ quantile of $\tilde{\pi}(p)$, 
we specify a sequence of increasing  success probability values $0 < p_1 < p_2 < \dots < p_N < 1$.
For $n = 1, \cdots, N$ we consecutively test  $H^{LE}_0 (q_0 ; p_n)$ at level $\alpha$ until the null hypothesis is accepted or all the null hypotheses have been rejected.
If $H^{LE}_0 ( q_0 ; p_1)$ is accepted, the confidence interval is  $[ 0, 1]$;
otherwise, it is $[q_{ll},1]$,  where $q_{ll}$ is the largest $p_n$ for which $H^{GE}_0 ( q_0 ; p_n)$  was rejected.

\medskip
\subsection{Two-sided confidence intervals}
 $1 - \alpha$ two-sided confidence intervals are intersections of the corresponding
right-tailed and left-tailed $1 - \alpha/2$ confidence intervals.
e.g.,  if  $[0, q_{ul}]$ is a left-tailed $1 - \alpha/2$ confidence interval for $CDF_{{\pi}} (p_0)$ and 
 $[q_{ll}, 1 ]$ is a right-tailed  $1 - \alpha/2$ confidence interval for $CDF_{{\pi}} (p_0)$,
 then  $CI^{TS} = [q_{ll}, q_{ul} ]$ is a two-sided $1 - \alpha$ confidence interval for $CDF_{{\pi}} (p_0)$.
 Thus we get
 \begin{eqnarray*}
\lefteqn{ \Pr_{\vec{ X}}  \{  \  CDF_{{\pi}} (p_0) \notin [q_{ll}, q_{ul} ]  \}
  =   \Pr_{\vec{ X}}  \{  \  CDF_{{\pi}} (p_0)  \notin [q_{ll},1 ]   \; \vee \;   CDF_{{\pi}} (p_0)  \notin [0, q_{ul} ]  \} } \\
& \le &  \Pr_{\vec{ X}}  \{  \  CDF_{{\pi}} (p_0)  \notin [q_{ll},1 ]  \} 
+  \Pr_{\vec{ X}}  \{  \   CDF_{{\pi}} (p_0)  \notin [0, q_{ul} ]  \} \;  \le \;   \alpha / 2+ \alpha/2   \; = \;  \alpha.
 \end{eqnarray*}


\bigskip
\subsection{Deconvolution-based test for $H^{LE}_0$}

For the deconvolution-based test of $H^{LE}_0 (q_0 ; p_0) :   q_0 \le CDF_{\tilde{\pi}} (p_0)$,
the test statistics are also the posterior median of the CDF of the mixing distribution evaluated at the endpoints of the dyadic partition,
\[
\widehat{CDF} (a_j ; \vec{X}) = \opmedian_{\vec{\pi}_L | \vec{X}}  \;  \{  CDF_{\vec{\pi}_L} (a_j) \}.
\]
However, as  in this case the worst-case null mixing distribution is $\tilde{\pi} ^{max} (p ; q_0, p_0)$,
which assigns probability $q_0$ to the event $P = p_0$ and probability $1 - q_0$ to the event $P = 1$.
To construct smaller confidence intervals, we incorporate the shift parameter to allow us to consider the test statistic at larger success probabilities than $p_0$.
For $\rho \ge 0$, we define $p^*_j (a_j, \rho) = \oplogit^{-1} (a_j - \rho)$, and we reject null hypothesis  $H^{LE}_0 (q_0 ; p^*_j)$
for small values of test statistic $\widehat{CDF} (a_j ; \vec{X})$.
The significance level is evaluated by Monte Carlo simulation  
of the test statistic distribution under the worst-case mixing distribution corresponding to $H^{LE}_0 (q_0 ; p^*_j)$,
 \begin{equation} \label{dfn-pval-LT-b}
\oppvalue  ( a_j , \rho, \vec{x} )   = Pr_{\vec{X} \sim \tilde{\pi}^{max} (p ; \ q_0, p^*_j ) }  \left(   \ \widehat{CDF}( a_j ; \vec{x}) \  \ge  \ \widehat{CDF}( a_j ; \vec{X} )  \ \right),
\end{equation}
Per construction, if $H^{LE}_0 (q_0 ; p^*_j)$ is true
then $\tilde{\pi}^{max} (p ; q_0, p^*_j)$ is stochastically greater than $\tilde{\pi} (p)$, the mixing distribution that generated the data.
Thus, as in Proposition \ref{prop66}, Proposition \ref{prop11} implies that the test, reject reject $H^{LE}_0 (q_0 ; p^*_j)$ if the p-value in (\ref{dfn-pval-LT-b}) is less than 
or equal to $\alpha$, has significance level $\alpha$.

\bigskip
\bigskip
\subsection{Confidence curves for the mixing distribution} \label{const-curves}

To construct $1 - \alpha$ confidence curves for the mixing distribution we test null hypotheses $H^{LE}_0 (q_0, p_0)$
and  $H^{GE}_0 (q_0, p_0)$  for a grid of $q_0$ and $p_0$ values.
The upper confidence curve is produced by $1 - \alpha/2$ left-tailed confidence intervals for the CDF of the mixing distribution for each value of $p_0$.
The lower confidence curve is produced by $1 - \alpha/2$ right-tailed confidence intervals for the CDF of the mixing distribution for each value of $p_0$.

\medskip
We provide an algorithm for constructing the lower confidence curve with a given shift parameter value $\rho$.
The algorithm for constructing the upper confidence curve follows through the symmetry of the problem. 
We begin by specifying an  increasing sequence of success probability values $0 < p_1 < \cdots < p_{N_1} < 1$,
where we only consider success probability values that equal $\oplogit^{-1} (  a_j + \rho)$ for $j \in \{ 0, 1, \cdots, 2^L \}$,
 and an increasing  sequence of CDF values $0 < q_1 < \cdots < q_{N_2} < 1$.
For $j = 1, \cdots N_1$, we construct a right-tailed confidence interval for $CDF_{\tilde{\pi}} ( p_j)$, by consecutively testing 
$H^{GE}_0 (q_i, p_j)$ for $i = 1, \cdots, N_2$ until the null hypothesis is accepted or $i = N_2$.

Note that as rejecting $H^{GE}_0 (q_i, p_j)$ implies rejecting $H^{GE}_0 (q_i, p_l)$ for $j < l$,
then if the lower limit of the confidence interval  $CDF_{\tilde{\pi}} ( p_{j'})$ is $q_{i'}$ for $i' \in \{ 1, \cdots, N_2-1 \}$,
to construct the confidence interval for $CDF_{\tilde{\pi}} ( p_{j'+1})$, we only need to consider testing 
$H^{GE}_0 (q_i, p_{j'+1})$ for $i = i'+1, \cdots, N_2$.
Furthermore, if the confidence interval  $CDF_{\tilde{\pi}} ( p_{j'})$ is $[q_{N_2},1]$,
then it is also the confidence interval for  $CDF_{\tilde{\pi}} ( p_{j})$  for all $j > j'$.

\medskip
We use a data driven procedure for determining the shift parameter value for constructing the confidence curves.
The procedure is similar to Algorithm \ref{algo3}, the only difference is that in Step 3, 
instead of selecting $\rho$ that minimizes the length of the confidence interval for the $q_0$ quantile of the mixing distribution, 
the procedure selects the value of $\rho$ that minimizes the sum of lengths of the nine calibration data confidence intervals
for the CDF of the mixing distribution at the success probabilities nearest to $0.1, 0.2, \cdots, 0.9$.


\bigskip
\bigskip

\section{Comparison of shift parameter values for the Efron (2016) data and for a simulated example} \label{subsec:appendix_Additional_Efron}

\begin{figure}[htbp] 
	\centering
	\includegraphics[width=.8\textwidth]{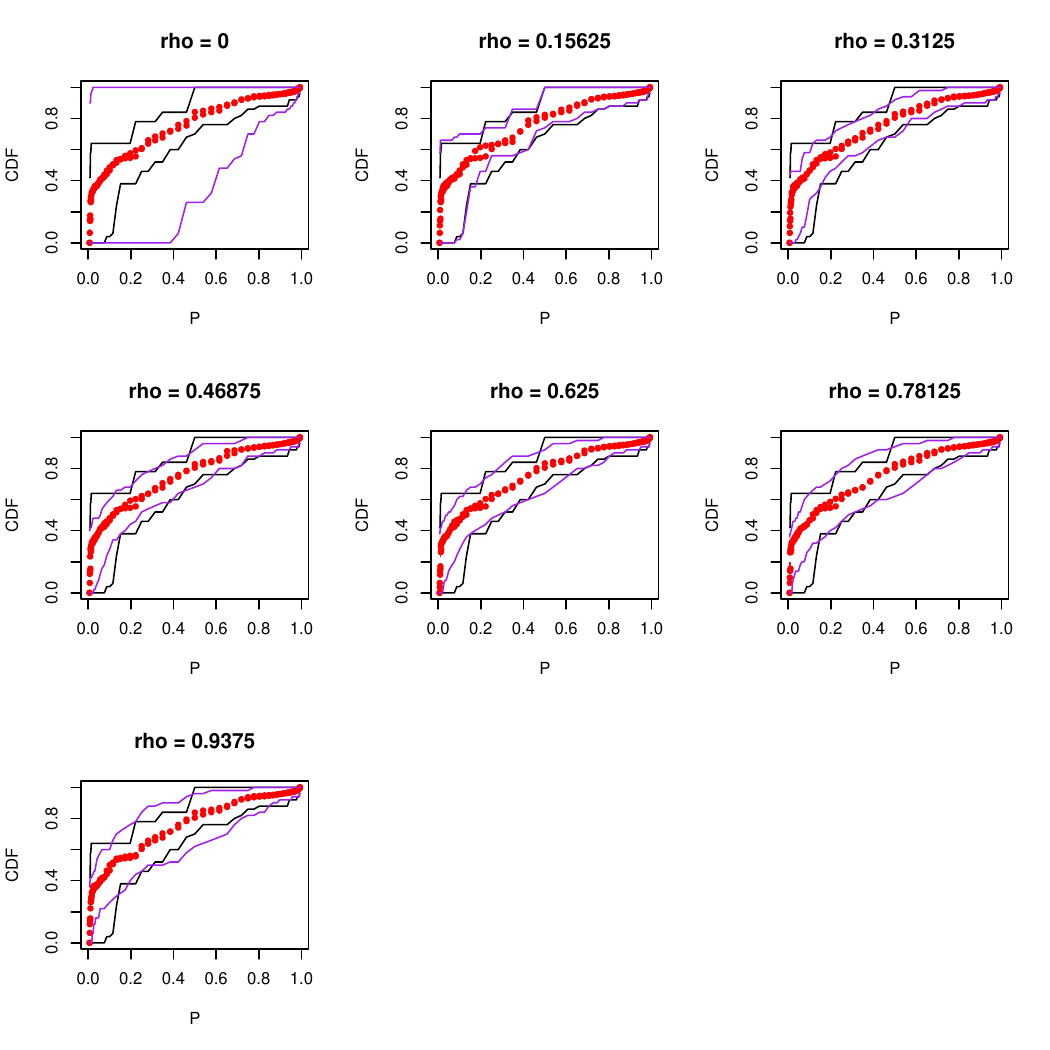}
	\caption{Comparison of the confidence curves produced by our data driven procedure for the analysis of the Efron (2016) data 
	(black curves) to alternative analyses with fixed values of $\rho$ that use all the data for constructing the confidence curves (purple curves). 
	The red points are the posterior medians of the mixing distribution.} 
	\label{figure:2a_Efron_fixed_rho_comparisons_holdout_0_2}
\end{figure}

For the analysis of the Efron (2016) data in Example \ref{exa0} 
we applied a $6$ level FPT model on endpoints vector $\vec{a} = (a_0, \cdots a_{64})$, with $a_j =  j \cdot 10 / 64 - 5$.
We used the data driven procedure with 20\% calibration data for determining $\rho$.
The value of $\rho$ selected by our procedure was $0.15625$.  
Figure \ref{figure:2a_Efron_fixed_rho_comparisons_holdout_0_2} 
compares the confidence curves shown in Example~\ref{exa0}, 
to alternative analysis, where all data is used for constructing the confidence curves with seven fixed values of $\rho$. 
We see that for $\rho = 0.15625$ the confidence curves obtained with adaptive selection of $\rho$ are similar to the ones obtained when all data is used to analysis.
For  $\rho = 0$ the confidence curves are very wide, $\rho = 0.3125$ seems to yield the tightest confidence curves,
and that larger $\rho$ values produce confidence curves that are tighter for small success probabilities and wider for large success proabibilities.
 
\begin{figure}[htbp] 
\centering
\includegraphics[width=.8\textwidth]{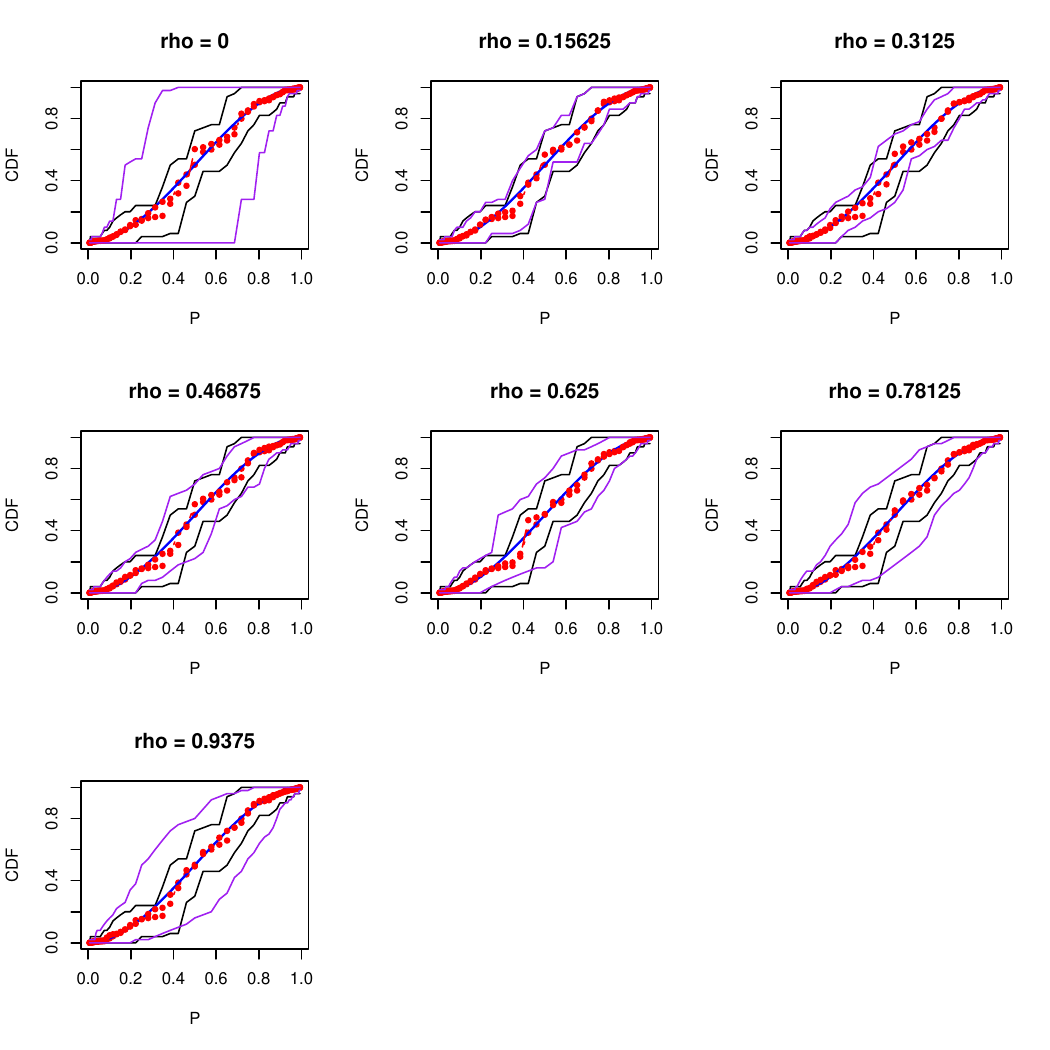}
	\caption{Comparison of the confidence curves produced by our data driven procedure 
	(black curves) to alternative analyses with fixed values of $\rho$ that use all the data for constructing the confidence curves (purple curves)
	on simulated data.
	The red points are the posterior medians of the mixing distribution.} 
\label{figure:3a_Calibration_1_fixed_rho_comparisons}
\end{figure}

 Figure \ref{figure:3a_Calibration_1_fixed_rho_comparisons} compares the same modelling setup on
 simulated data consisting of $K=1000$ binomial count, with sample sizes $m_k = 20$, for success probabilities sampled from the $Beta (2,2)$ mixing distribution.
Also in this example, the  value of $\rho$ selected by our procedure was $0.15625$.  
We see that $\rho = 0$ yields very wide confidence curves,
values of $\rho$  of $0.15625$ and $0.3125$  yield tight confidence curves,
and that larger $\rho$ values produce wider confidence curves.

\end{document}